\documentclass[]{valhalla}

\usepackage{amsmath}
\let\CODesignOriginalEqref\eqref

\usepackage{amsmath,amsfonts,bm}

\def\eqref#1{equation~\ref{#1}}

\def\1{\bm{1}}

\DeclareMathAlphabet{\mathsfit}{\encodingdefault}{\sfdefault}{m}{sl}
\SetMathAlphabet{\mathsfit}{bold}{\encodingdefault}{\sfdefault}{bx}{n}

\let\eqref\CODesignOriginalEqref

\usepackage{iftex}
\usepackage{url}
\usepackage{booktabs}
\usepackage{graphicx}
\usepackage{float}
\usepackage{xcolor}
\usepackage{colortbl}
\usepackage{makecell}
\usepackage{pifont}
\usepackage{wrapfig}
\usepackage{hyperref}
\usepackage{array}
\usepackage{multirow}
\usepackage{amsmath}
\usepackage{amssymb}
\usepackage{amsfonts}
\usepackage{nicefrac}
\usepackage{microtype}
\usepackage{longtable}
\usepackage{enumitem}
\usepackage{subcaption}
\usepackage{cleveref}
\usepackage[toc,page,header]{appendix}
\usepackage{minitoc}
\usepackage{tcolorbox}
\tcbuselibrary{most}

\makeatletter
\newcommand{\EpibenchAppendixHeadingSpacing}{%
    \renewcommand{\section}{\@startsection{section}{1}{\z@}{-2.5ex plus -0.4ex minus -0.2ex}{1.1ex plus 0.2ex}{\large\sffamily\bfseries\color{valhallaprimary}\raggedright}}%
    \renewcommand{\subsection}{\@startsection{subsection}{2}{\z@}{-1.8ex plus -0.3ex minus -0.1ex}{0.65ex plus 0.1ex}{\normalsize\sffamily\bfseries\color{valhallaprimary}\raggedright}}%
    \renewcommand{\subsubsection}{\@startsection{subsubsection}{3}{\z@}{-1.3ex plus -0.2ex minus -0.1ex}{0.5ex plus 0.1ex}{\normalsize\sffamily\bfseries\color{valhallaprimary}\raggedright}}%
}
\newcommand{\EpibenchNeedspace}[1]{%
    \par\begingroup
    \dimen@=#1\relax
    \dimen@ii=\pagegoal
    \advance\dimen@ii by -\pagetotal
    \ifdim\dimen@>\dimen@ii
        \pagebreak
    \fi
    \endgroup
}
\makeatother

\newcounter{epialgorithm}

\definecolor{ReasoningBand}{HTML}{F1EAFE}
\definecolor{NonreasoningBand}{HTML}{F7F3EE}
\newsavebox{\exampleboxcontent}
\newsavebox{\systempromptboxcontent}
\newsavebox{\caseboxcontent}
\newlength{\exampleboxwidth}
\newlength{\systempromptboxwidth}
\newlength{\caseboxwidth}

\renewcommand{\sectionfont}{\fontsize{12}{14.5}\selectfont}
\renewcommand{\subsectionfont}{\fontsize{11}{13}\selectfont}
\titleformat*{\section}{\sectionfont\sffamily\bfseries\color{valhallaprimary}\raggedright}
\titleformat*{\subsection}{\subsectionfont\sffamily\bfseries\color{valhallaprimary}\raggedright}
\titleformat*{\subsubsection}{\normalsize\sffamily\bfseries\color{valhallaprimary}\raggedright}
\titlespacing*{\section}{0pt}{2.5ex plus .4ex minus .2ex}{1.1ex plus .2ex}
\titlespacing*{\subsection}{0pt}{1.8ex plus .3ex minus .1ex}{.65ex plus .1ex}
\titlespacing*{\subsubsection}{0pt}{1.3ex plus .2ex minus .1ex}{.5ex plus .1ex}

\AtBeginEnvironment{thebibliography}{\interlinepenalty=10000}

\newenvironment{CODesignKeepParagraph}{\par\begingroup\interlinepenalty=10000}{\par\endgroup}
\newenvironment{CODesignKeepBlock}{\par\begin{samepage}}{\par\end{samepage}}

\renewcommand{\authorfont}{\fontsize{8.8}{12}\selectfont}
\fancypagestyle{firststyle}{%
  \fancyhf{}%
  \fancyhead[L]{%
    \vskip 2.6mm%
    \includegraphics[height=10mm]{\paperlogopath}%
  }%
  \fancyhead[R]{%
    \vskip 2.6mm%
    \includegraphics[height=10mm]{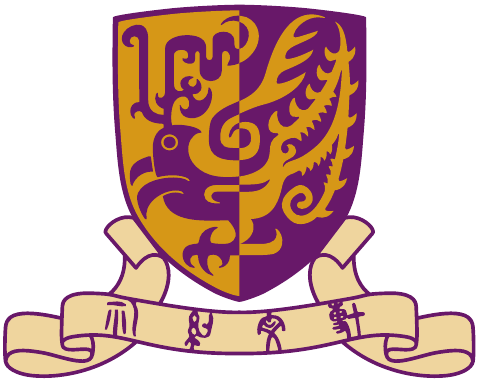}\hspace{4mm}%
    \raisebox{1mm}{\includegraphics[height=8mm]{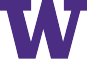}}\hspace{4mm}%
    \includegraphics[height=10mm]{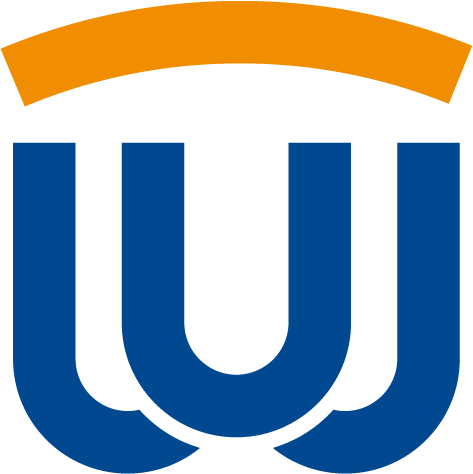}\hspace{4mm}%
    \includegraphics[height=10mm]{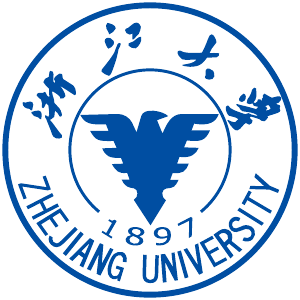}%
  }%
  \fancyfoot[C]{\thepage}%
}

\usepackage{algorithm}
\usepackage{algpseudocode}

\paperlogo{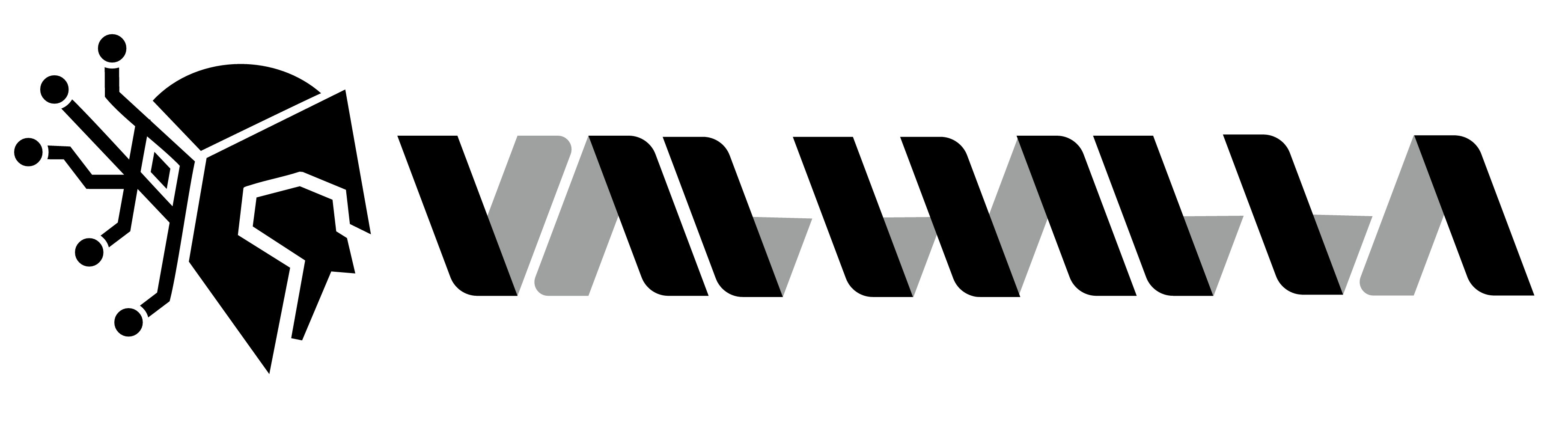}
\papertagline{}
\title{CODesign: Consistency from Data to Trajectory in All-Atom Protein Binder Co-Design}

\author[1,2,*]{Yuanle Mo}
\author[3,*]{Bo Qiang}
\author[4,*]{Haitao Lin}
\author[1,5]{Qinghan Wang}
\author[1,2]{Gang Du}
\author[1,2,\dagger]{Odin Zhang}
\author[2,\dagger]{Pheng Ann Heng}

\affiliation[1]{Valhalla Technology}
\affiliation[2]{The Chinese University of Hong Kong}
\affiliation[3]{University of Washington}
\affiliation[4]{Westlake University}
\affiliation[5]{Zhejiang University}
\contribution[*]{Equal Contribution}
\contribution[\dagger]{Corresponding Authors}
\correspondence{\email{odin@valtech.ai}; \email{pheng@cse.cuhk.edu.hk}}

\abstract{
The central challenge in \textit{de novo} protein design is generating plausible, mutually compatible structures and sequences, such that each designed sequence folds into its intended structure and the structure accommodates that sequence. Compared to typical two-stage design methods, which decouple the modeling of the interdependent modalities, co-design models improve the cross-modal consistency by jointly generating sequences and structures.
However, naively generating sequences and structures simultaneously does not ensure their consistency. To address this challenge, we propose CODesign framework. We improve data consistency by generating approximately 105,000 consistency-distilled dimers. We further promote consistency through a multimodal joint flow model that captures the joint distribution of sequences, backbone structures, and local atomic configurations, together with a consistency-aware joint resampling strategy that iteratively refines sequences and side chains. Experiments show that CODesign achieves state-of-the-art performance with the highest in silico success rates on both protein- and ligand-target binder design. Ablation studies also demonstrate our distilled dataset increases performance by 70.9\%, which can be further improved by our proposed resampling mechanism with negligible additional computational cost. Code, model weights and the new dataset will be completely open-source.
}

\begin{document}
\maketitle

\begin{figure}[h]
\centering
\includegraphics[width=0.99\linewidth]{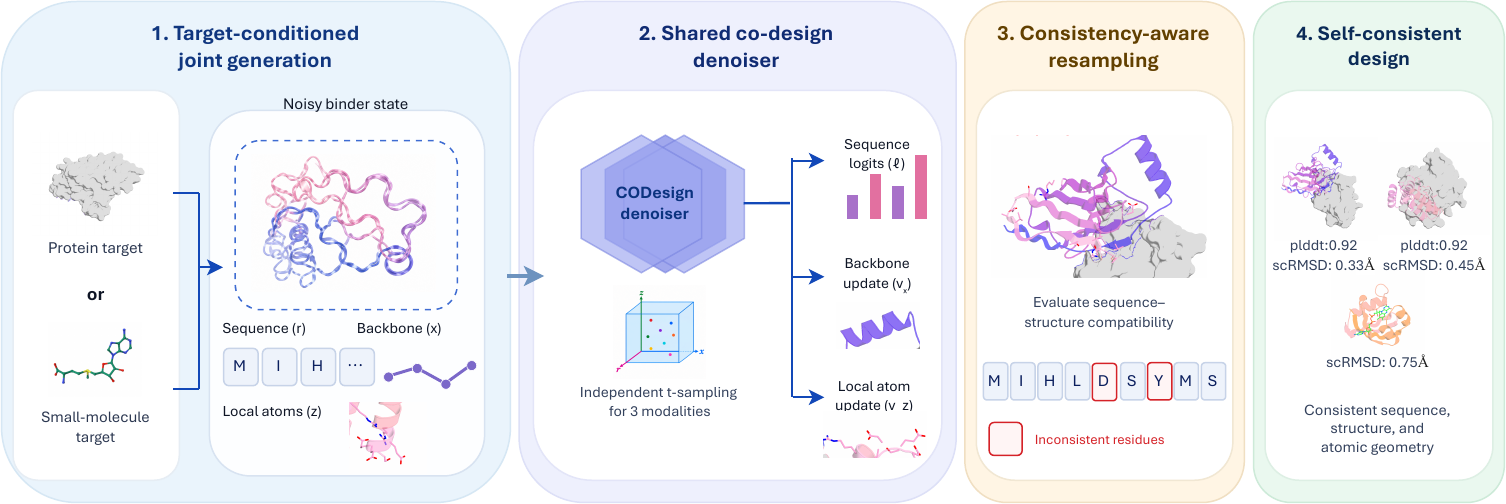}
\caption{CODesign jointly generates sequence, backbone, and local atomic states conditioned on either a protein or small-molecule target. A shared co-design denoiser is trained with independent modality-time sampling and predicts sequence logits, backbone updates, and local-atom updates. During generation, we apply consistency-aware joint resampling for residue and side-chain refining. This overall framework generates protein binders with consistent all-atom structures.}
\label{fig:abstract_fig}
\end{figure}

\section{Introduction}
\label{sec:introduction}

\textit{De novo} protein design offers a promising avenue for developing a wide range of therapeutics \citep{watson2023novo}. Protein design models can be formulated as probabilistic models of protein sequence--structure pairs conditioned on a given target. To capture biophysical interactions under geometric constraints, prior methods typically adopt a two-stage approach, sequentially generating protein backbones and inverse-folding them into candidate sequences~\citep{dauparas2022robust}. 
Although two-stage methods have achieved strong results in monomer and enzyme
design, their low pass rates on challenging targets make them inefficient ~\citep{watson2023novo, ahern2026atom}.

Recent co-design methods jointly generate discrete sequences and structures, 
which could model residue-specific interactions~\citep{butcher2025novo}.
Despite their increased expressiveness, they still underperform two-stage methods or generate a biased residue-type distribution. 
Latent diffusion offers an alternative by jointly generating continuous sequence embeddings and backbone coordinates~\citep{geffner2026proteina, didi2026scaling}, but does not explicitly model discrete residue identities and thus falls outside explicit sequence--structure co-design.

Despite these advances, ensuring sequence--structure consistency remains a central challenge for binder co-design. We highlight two key bottlenecks: \textit{(1) the scarcity of self-consistent training data for protein complexes} and \textit{(2) the difficulty of maintaining sequence--structure compatibility throughout sampling}.
Complexes in the Protein Data Bank (PDB) remain limited in number and structural diversity~\citep{didi2026scaling}. Although predicted complexes can expand the training set, many synthetic sequence--structure pairs exhibit poor self-consistency: refolding their sequences does not reliably recover the paired structures. \citet{reidenbach2025consistent} identified a similar limitation in predicted monomers from the widely used AlphaFold Database (AFDB)~\citep{varadi2022alphafold,han2026alphafold}. 
Besides, residue identities selected from partially resolved structures may become incompatible with the evolving geometry~\citep{rector2026general} during sampling. 
Existing methods often rely on auxiliary sequence-revision modules or external structure-prediction guidance, providing indirect consistency signals and potentially incurring substantial computational overhead~\citep{rector2026general,cheng2026code,didi2026scaling}. This motivates integrating compatibility assessment and correction directly into joint sequence--structure sampling.

\looseness=-1
To address these bottlenecks, we introduce \textbf{CODesign}, a consistency-driven multimodal flow framework that extends Prote\'{i}na-Atom\'{i}stica~\citep{reidenbach2025consistent} from monomer generation to target-conditioned all-atom protein binder co-design.
\textit{For improving training-data consistency,}
we construct \textbf{104.5k} distilled sequence--structure pairs by jointly redesigning both partners of experimental and predicted complexes with ProteinMPNN~\citep{dauparas2022robust,reidenbach2025consistent} and co-folding the redesigned sequences with ESMFold2-Fast. Among high-confidence candidates, we select the structure that best recovers the reference complex, 
emphasizing interface qualities.
\textit{For maintaining consistency during sampling,} we propose {Consistency-Aware joint Resampling (\textbf{CAR})}, which exploits CODesign's intrinsic inverse-folding capability enabled by training with decoupled modality times and structural conditioning.
CAR reassesses residue choices against the evolving backbone, selectively resampling incompatible identities together with their local atomic states. These corrections also enter self-conditioning, modifying subsequent backbone vector fields so that $C_\alpha$ trajectories can adapt without direct coordinate resets. CAR thus couples sequence revision with all-atom refinement, without an external inverse-folding model or folding-based reward oracle.

\looseness=-1
We evaluate CODesign on protein- and ligand-target binder design benchmarks \citep{didi2026scaling}. Experiments demonstrate that our method achieves state-of-the-art performance with the highest in silico unique success rates among evaluated baselines. Ablation studies show that our distilled dataset can increase unique successes by a substantial margin of \textbf{70.9\%}, which can be further improved by 4\% using the proposed resampling algorithm. These results underscore the potential of CODesign as a powerful next-generation tool for developing novel protein binders.



\section{Related Work}
\begin{CODesignKeepParagraph}
Protein generative modeling has progressed from backbone generation followed by inverse folding in RFdiffusion \citep{watson2023novo,dauparas2022robust} toward joint sequence--structure modeling. In discrete all-atom design, residue identity determines atomic composition. Atom14/37 provides a fixed-dimensional representation using 14/37 masked atom slots per residue, enabling side-chain interaction modeling~\citep{qu2024p,butcher2025novo, reidenbach2025consistent}. A complementary family of methods avoids explicit residue identities during generation by jointly modeling structures and continuous sequence embeddings \citep{geffner2025proteina,geffner2026proteina,didi2026scaling}. Building on these paradigms, task-specific models extend co-design to target-conditioned binder design using objectives or guidance tailored to intermolecular interfaces \citep{cheng2026code,didi2026scaling}. 
\end{CODesignKeepParagraph}


\section{Methods}
\label{sec:codes}


\paragraph{Consistency in target-conditioned co-design.}
Let $\mathbf{c}^{\mathrm{target}}$ denote the representation of a protein or small-molecule target. We represent a binder by its amino-acid sequence $\mathbf{r}$, $C_\alpha$ coordinates $\mathbf{x}$, and non-$C_\alpha$ coordinates $\mathbf{z}$. Target-conditioned co-design seeks to learn the conditional joint distribution
\begin{equation}
p^\star(\mathbf{r},\mathbf{x},\mathbf{z}
\mid \mathbf{c}^{\mathrm{target}}).
\label{eq:conditional_joint}
\end{equation}
This distribution specifies the desired endpoint but does not impose a particular generation order. In CODesign, the three modalities evolve within a shared sampling trajectory, allowing sequence and structure predictions to influence one another as generation proceeds.

A consistent sample must satisfy two complementary conditions. First, its sequence should be compatible with the target-conditioned backbone geometry. Second, its complete all-atom structure should be recoverable from the sequence in the presence of the target. We characterize these conditional relationships as
\begin{equation}
\mathcal{C}(
\mathbf{r},\mathbf{x},\mathbf{z};
\mathbf{c}^{\mathrm{target}})
=
\underbrace{
\log p^\star(
\mathbf{r}\mid
\mathbf{x},\mathbf{c}^{\mathrm{target}})
}_{\mathcal{C}_{\mathrm{seq}}}
+
\lambda
\underbrace{
\log p^\star(
\mathbf{x},\mathbf{z}\mid
\mathbf{r},\mathbf{c}^{\mathrm{target}})
}_{\mathcal{C}_{\mathrm{str}}},
\label{eq:consistency}
\end{equation}
where $\mathcal{C}_{\mathrm{seq}}$ measures sequence compatibility using residue-agnostic $C_\alpha$ geometry, $\mathcal{C}_{\mathrm{str}}$ measures all-atom structural recoverability from the sequence, and $\lambda$ balances the two directions. 
Low-consistency training pairs provide conflicting supervision for learning the target distribution in Equation~\ref{eq:conditional_joint}, while initially compatible modalities may still diverge during the sampling trajectory.

CODesign addresses these two failure modes within a unified framework. At the data level, \textsc{Codimer} uses inverse folding and co-folding as surrogates for the two conditional terms in Equation~\ref{eq:consistency}, producing more consistent training pairs (Section~\ref{sec:codimer}). At the model level, independently sampled modality times enable the multimodal flow to learn both structure-to-sequence and sequence-to-structure conditional regimes (Section~\ref{sec:codes-model}). At the sampling level, consistency-aware joint resampling uses the learned conditionals to identify incompatible residue choices and jointly update their identities and local atomic states (Section~\ref{sec:codes-cajr}).

\subsection{\textsc{Codimer}: Consistency-Distilled Protein Dimers}
\label{sec:codimer}

\paragraph{Limitations of existing complex data.}
Existing approaches either pretrain on predicted AFDB monomers \citep{cheng2026code}, or construct pseudo-complexes from domain--domain interactions in predicted monomers \citep{didi2026scaling}.
Nonetheless, they do not explicitly ensure that the paired sequence has high probability under $\mathcal{C}_{\mathrm{seq}}$. 
Predicted structures may be individually plausible while failing to recover the same complex geometry when refolded. This issue is observed in high-confidence AFDB heterodimers \citep{han2026alphafold}: only $8.49\%$ are recovered by folding oracles within a complex-level $C_\alpha$ scRMSD of $2$\,\AA. 

\paragraph{Consistency distillation.}
We construct \textsc{Codimer} by approximating the two conditional terms in Equation~\ref{eq:consistency} through inverse folding and co-folding. We use ESMFold2-Fast for its computational efficiency and the demonstrated correlation of its predictions with experimental binder success~\citep{claudescience2026autonomous}. Given a reference complex structure $\mathbf{x}_0$, ProteinMPNN jointly redesigns both chains to produce $K=4$ candidate sequence pairs, which ESMFold2-Fast then co-folds into the corresponding all-atom complexes:
\begin{equation}
\mathbf{r}_k
\sim q_{\mathrm{MPNN}}(\mathbf{r}\mid\mathbf{x}_0),
\hat{\mathbf{x}}_k
\sim q_{\mathrm{fold}}(\mathbf{x}\mid\mathbf{r}_k),
\qquad k=1,\ldots,K.
\label{eq:distillation}
\end{equation}
The first step promotes sequence compatibility with the reference geometry, while the second evaluates whether the redesigned sequence recovers that geometry. We retain candidates that satisfy
\begin{equation}
\overline{\mathrm{pLDDT}}(\hat{\mathbf{x}}_k) \geq 0.7,
\qquad
\operatorname{ipAE}(\hat{\mathbf{x}}_k) \leq 10\,\text{\AA}.
\label{eq:confidence_filters}
\end{equation}
Among the qualifying candidates, we select the one with the lowest complex-level $C_\alpha$ RMSD to the reference structure:
\begin{equation}
k^\star
=
\underset{k \in \mathcal{A}}{\arg\min}\;
d_{C_\alpha}(\hat{\mathbf{x}}_k,\mathbf{x}_0),
\label{eq:selection}
\end{equation}
where $\mathcal{A}$ denotes the set of candidates satisfying both confidence criteria.
We apply this procedure to AFDB high-confidence heterodimers \citep{han2026alphafold}, HumanPPI predictions \citep{zhang2025predicting}, and experimentally resolved PINDER holo complexes \citep{kovtun2024pinder}. After source-specific filtering and restricting each chain to 40--300 residues, 191,836 reference complexes are processed to produce 104,528 distilled sequence--structure pairs. Each retained structure is co-folded directly from its paired sequence, yielding supervision that better approximates the consistent joint distribution in Equation~\ref{eq:conditional_joint}. Further construction details are provided in Appendix~\ref{app:codimer}.

\begin{figure}[t]
\centering
\includegraphics[width=\linewidth]{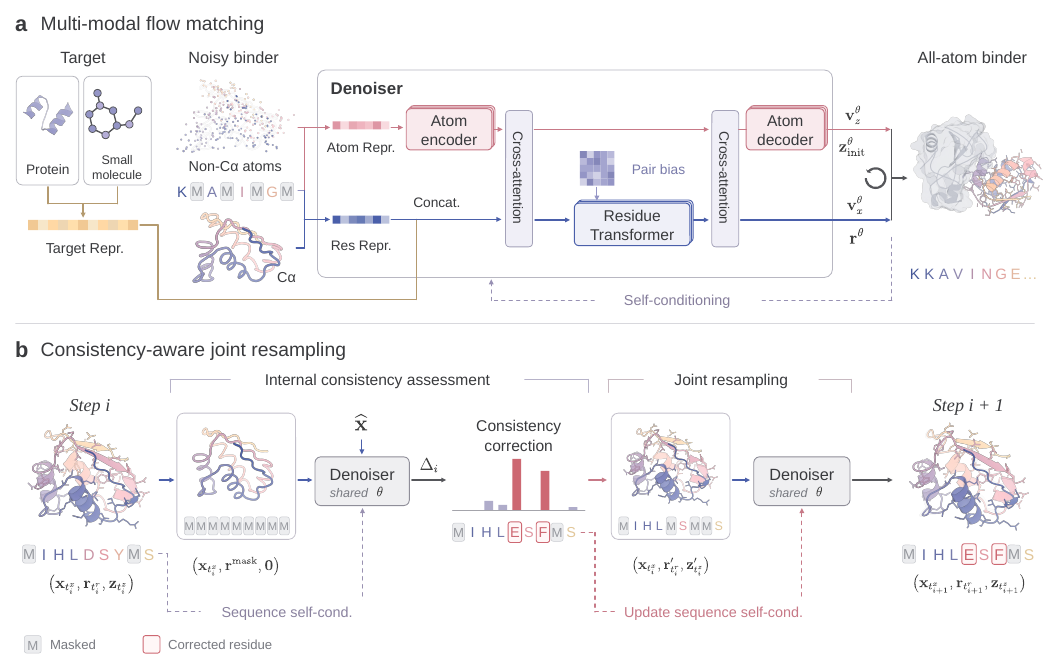}
\caption{\textbf{Overview of CODesign and CAR.} \textbf{(a)} CODesign jointly generates a binder sequence and all-atom structure conditioned on a protein or small-molecule target using coupled residue- and atom-level denoising with self-conditioning. \textbf{(b)} CAR evaluates decoded residues against the evolving backbone, selects incompatible identities using the model’s conditional sequence predictions, and jointly resamples their sequence and local atomic states.}
\label{fig:overview}
\end{figure}

\subsection{Multi-Modal Flow Matching}
\label{sec:codes-model}


\paragraph{Target-conditioned explicit joint flow.}
A target-conditioned co-design model should approximate $p^\star(\mathbf{r},\mathbf{x},\mathbf{z}\mid\mathbf{c}^{\mathrm{target}})$ while evolving sequence, backbone, and local atomic geometry within a shared sampling trajectory. This coupling allows sequence and structure predictions to influence one another throughout sampling. Because the three modalities can resolve at different stages of the trajectory, we independently sample their modality times $t_x,t_r,t_z\in[0,1]$ during training, exposing the model to different conditional regimes. A nearly clean backbone paired with a highly masked sequence induces the structure-to-sequence prediction underlying $\mathcal{C}_{\mathrm{seq}}$, whereas a nearly clean sequence paired with noisy coordinates induces the sequence-to-structure prediction underlying $\mathcal{C}_{\mathrm{str}}$.

\paragraph{Fixed-dimensional all-atom representation.}
To jointly evolve sequences and all-atom structures in a fixed-dimensional state, we use Atom37, which assigns slots to the 37 possible heavy-atom positions. For a binder of length $n$, we represent the sequence as $\mathbf{r}\in\{0,\ldots,19\}^n$, the $C_\alpha$ coordinates as $\mathbf{x}\in\mathbb{R}^{n\times3}$, and the non-$C_\alpha$ coordinates as $\mathbf{z}\in\mathbb{R}^{n\times36\times3}$, where $\mathbf{z}_{ij}$ is the position of atom slot $j$ relative to the $C_\alpha$ atom of residue $i$. Residue-specific validity masks exclude nonexistent atoms. For masked residues, we also hide the non-$C_\alpha$ coordinates and atom occupancy to prevent the atomic representation from revealing residue identity.

\paragraph{Multimodal conditional paths.}
To realize the conditional regimes described above, we define modality-specific paths from corrupted states at $t=0$ to clean data at $t=1$. We use continuous flow matching for the coordinates and discrete flow matching for residues~\citep{campbell2024generative}:
\begin{align}
\mathbf{x}_{t_x}
&=
t_x\mathbf{x}
+
(1-t_x)\mathbf{x}_0,
\label{eq:codes-flow-x}\\
p_{t_r}(\,\cdot\mid r_i)
&=
t_r\delta_{r_i}
+
(1-t_r)\delta_{\mathtt{m}},
\qquad
r_{t_r,i}\sim p_{t_r}(\,\cdot\mid r_i),
\label{eq:codes-flow-r}\\
\mathbf{z}_{t_z}
&=
t_z\mathbf{z}
+
(1-t_z)\mathbf{z}_0.
\label{eq:codes-flow-z}
\end{align}
Here, $\mathbf{x}_0$ and $\mathbf{z}_0$ denote the transformed backbone and local-coordinate reference endpoints, respectively; $\delta_a$ denotes a point mass at token $a$; and $\mathtt{m}$ is the mask token. Thus, residue $i$ retains its clean identity with probability $t_r$ and is masked otherwise. 
The backbone reference endpoint uses centered Gaussian noise with a randomized three-dimensional center, requiring the model to infer binder placement relative to the target. Exact noise transformations and modality-time distributions are provided in Appendix~\ref{app:codes-noise}.

\paragraph{Target-conditioned denoiser.}
The overall framework of CODesign is illustrated in Figure~\ref{fig:overview}a. Our network, parameterized by $\theta$, couples a residue-level Transformer trunk with an atom-level Transformer encoder--decoder~\citep{reidenbach2025consistent}. Target embeddings are combined with noisy binder residue embeddings and processed using pair-biased attention over geometric features within and between the binder and target~\citep{didi2026scaling}. Cross-attention exchanges information between the residue and atom trunks, connecting target context and global backbone geometry to local atomic structure.

Given $\mathbf{x}_{t_x}$, $\mathbf{r}_{t_r}$, $\mathbf{z}_{t_z}$, $\mathbf{c}^{\mathrm{target}}$, and $t_x,t_r,t_z$, the network predicts a backbone vector field $\mathbf{v}_x^\theta$, sequence logits $\boldsymbol{\ell}^\theta$, a non-$C_\alpha$ vector field $\mathbf{v}_z^\theta$, and pseudo-atom initialization outputs $\mathbf{b}^\theta$ for masked residues. Self-conditioning uses a detached preliminary prediction and provides features derived from predicted $C_\alpha$ coordinates and sequence identities.

To strengthen the two conditional relationships in Equation~\ref{eq:consistency}, we additionally provide clean-backbone or clean-sequence feature channels intermittently during training. These modes retain the independently sampled noisy states and modality times while strengthening the structure-to-sequence and sequence-to-structure predictions, respectively. They are applied with probability $0.1$ during protein training and $0.15$ during ligand adaptation, with further details provided in Appendix~\ref{app:codes-conditioning}.

\paragraph{Training objective.}
The denoiser jointly predicts the clean backbone, residue identities, and local atomic geometry. We reconstruct the clean backbone from its vector field as
\begin{equation}
\widehat{\mathbf{x}}
=
\mathbf{x}_{t_x}
+
(1-t_x)\mathbf{v}_x^\theta.
\label{eq:codes-clean-x}
\end{equation}
The backbone and sequence objectives are
\begin{equation}
\mathcal{L}_x
=
\operatorname{MSE}(\widehat{\mathbf{x}},\mathbf{x}),
\qquad
\mathcal{L}_r
=
\operatorname{CE}(\boldsymbol{\ell}^\theta,\mathbf{r}),
\label{eq:codes-loss-xr}
\end{equation}

For local atomic geometry, let
\begin{equation}
V_i
=
\mathbf{1}\{r_{t_r,i}\neq\mathtt{m}\}
\end{equation}
indicate whether residue $i$ is revealed, and let $M_{ia}\in\{0,1\}$ indicate whether non-$C_\alpha$ atom slot $a\in\{1,\ldots,36\}$ exists for its ground-truth identity. We denote by $\boldsymbol{\epsilon}^{\mathrm{raw}}_{z,ia}$ the stored Gaussian draw used to construct the local-coordinate reference endpoint. Revealed residues are supervised through $\mathbf{v}_z^\theta$, whereas masked residues are supervised through $\mathbf{b}^\theta$:
\begin{equation}
\mathcal{L}_z
=
\frac{1}{n}
\sum_{i,a}M_{ia}
\left\|
V_i\mathbf{v}_{z,ia}^\theta
+
(1-V_i)\mathbf{b}_{ia}^\theta
-
\left(
\mathbf{z}_{ia}
-
\boldsymbol{\epsilon}^{\mathrm{raw}}_{z,ia}
\right)
\right\|_2^2.
\label{eq:codes-loss-z}
\end{equation}
Thus, the same atomic target supervises coordinate refinement at revealed residues and pseudo-atom initialization at masked residues. Our core training objective is
\begin{equation}
\mathcal{L}
=
\mathcal{L}_x
+
\mathcal{L}_r
+
\mathcal{L}_z.
\label{eq:codes-loss-total}
\end{equation}
Exact normalization, noise construction, and auxiliary training objectives are provided in Appendix~\ref{app:codes-loss}.

\subsection{Consistency-Aware Joint Resampling}
\label{sec:codes-cajr}

Learning from consistent sequence--structure pairs does not prevent residue choices from becoming incompatible with the structure during sampling. We introduce Consistency-Aware joint Resampling (\textbf{CAR}), which uses the generator's own inverse-folding capability to identify such choices and jointly reset their sequence identities and local atomic states for resampling. CAR uses sequence--structure compatibility rather than random remasking to determine which residues to revisit, and, unlike search guided by external folding rewards~\citep{didi2026scaling}, obtains its correction signal directly from the denoiser within a single sampling trajectory.

\paragraph{Internal consistency assessment.}
Independent modality times during training allow the model to observe low-noise backbones ($t_x\approx1$) together with highly masked sequences ($t_r\approx0$). Predicting residue identities from these inputs closely resembles inverse folding. The additional clean-backbone conditioning features used during training further strengthen this capability while retaining the independently sampled noisy states and modality times. CAR uses this learned structure-to-sequence relationship to reassess decoded residue identities as the structure evolves during sampling.

To make this assessment more reliable, we perform it during the later stages of backbone denoising. As illustrated in Figure~\ref{fig:overview}b, we first compute the predicted clean $C_\alpha$ backbone
\[
    \widehat{\mathbf{x}}
    =
    \mathbf{x}_{t_x}+(1-t_x)\mathbf{v}_x^\theta
\]
from the current denoising prediction. Let $\mathcal{U}=\{i:r_{t_r,i}\ne\mathtt{m}\}$ denote the currently revealed positions. We remask these positions to obtain a fully masked sequence $\mathbf{r}^{\mathrm{mask}}$. We also replace their non-$C_\alpha$ inputs with the zero-coordinate pseudo-atom state $\mathbf{z}^{\mathrm{mask}}=\mathbf{0}$, preventing residue-specific atomic composition from revealing their identities. The predicted clean backbone $\widehat{\mathbf{x}}$ is supplied through the clean-backbone conditioning channel. Meanwhile, we retain the decoded identities in the sequence self-conditioning input $\mathbf{r}^{\mathrm{sc}}$, encouraging refinement of the current sequence rather than an unrelated prediction from the backbone alone. Because substantially different amino-acid sequences can fold into similar structures~\citep{koehler2023sequence}, retaining the decoded sequence also prevents large deviations from the current sequence trajectory.

With these inputs, an additional forward pass performs a single-step inverse-folding assessment and produces sequence logits
\begin{equation}
    \widetilde{\boldsymbol{\ell}}^\theta
    =
    \left[
    \operatorname{Denoiser}_\theta\!\left(
        \mathbf{x}_{t_x},
        \mathbf{r}^{\mathrm{mask}},
        \mathbf{z}^{\mathrm{mask}},
        \mathbf{c}^{\mathrm{target}},
        t_x,0,t_z,
        \widehat{\mathbf{x}},
        \mathbf{r}^{\mathrm{sc}}
    \right)
    \right]_{\mathrm{seq}}.
    \label{eq:codes-cajr-assessment}
\end{equation}
Let
$q_i(a)=\operatorname{softmax}(\widetilde{\boldsymbol{\ell}}_i^\theta)_a$
denote the predicted probability of amino-acid type $a$ at position $i$. For each $i\in\mathcal{U}$, we compare the most probable type with the current identity:
\begin{equation}
    a_i^\star
    =
    \operatorname*{arg\,max}_{a\in\{0,\ldots,19\}}q_i(a),
    \qquad
    \Delta_i
    =
    \log q_i(a_i^\star)-\log q_i(r_{t_r,i}).
    \label{eq:codes-cajr-gap}
\end{equation}
A larger $\Delta_i$ indicates a stronger preference for replacing the current identity with $a_i^\star$. Among positions satisfying $a_i^\star\ne r_{t_r,i}$, we select those with the largest gaps, up to $\lceil\rho|\mathcal{U}|\rceil$ residues. Here, $\rho\in[0,1]$ controls the resampling fraction, and $\mathcal{S}$ denotes the selected positions.

\paragraph{Joint resampling.}
For each selected position $i\in\mathcal{S}$, we use the preferred residue type $a_i^\star$ to guide resampling. Because changing a residue identity also changes its atomic composition, we reset its sequence and local atomic states together. Specifically, we remask the residue, replace its non-$C_\alpha$ input with the zero-coordinate pseudo-atom state, and update its sequence self-conditioning:
\begin{equation}
    r_{t_r,i}\leftarrow\mathtt{m},
    \qquad
    \mathbf{z}_{t_z,i}\leftarrow\mathbf{0},
    \qquad
    r_i^{\mathrm{sc}}\leftarrow a_i^\star,
    \qquad i\in\mathcal{S}.
    \label{eq:codes-cajr-reset}
\end{equation}
The $C_\alpha$ coordinates and unselected residue states are not reset. We then run the denoiser again with the updated states and self-conditioning at the original times $t_x,t_r,t_z$, and continue sampling using the new predictions. The discrete flow samples residue identities from the updated logits $\boldsymbol{\ell}^\theta$; the preferred types $a_i^\star$ guide this process through self-conditioning rather than being assigned directly. Newly revealed atoms are initialized by the pseudo-atom initialization output $\mathbf{b}^\theta$ and subsequently refined by $\mathbf{v}_z^\theta$.

Although $\mathbf{x}_{t_x}$ is not reset directly, the additional forward pass recomputes $\mathbf{v}_x^\theta$ using the revised sequence self-conditioning. The subsequent backbone update can therefore respond to the preferred residue types. CAR thus corrects the sequence while allowing both the backbone and local atomic geometry to adjust, rather than refining the sequence alone. Sampling schedules and implementation details are provided in Appendix~\ref{app:codes-sampling}.

\subsection{Stagewise Training}
\label{sec:codes-stages}

\begin{CODesignKeepParagraph}
We first train the protein-conditioned generator on Teddymer \citep{didi2026scaling} and experimental PDB data for 60k iterations with a batch size of 5, and then continue training for 10k iterations with \textsc{Codimer} incorporated into the mixture. The first stage enables the model to gain fundamental knowledge of protein-protein interactions from large-scale data, while the second introduces targeted consistency supervision. For small-molecule targets, we adapt the protein-conditioned model using LoRA~\citep{hu2022lora} to avoid overfitting, training on a mixture of PLINDER protein-ligand complexes~\citep{durairaj2024plinder} and monomers from Consistency Distilled Synthetic Protein Database (CDDB). The CDDB monomers retain consistency supervision alongside ligand-specific conditioning. We adopt a batch size of 3 for the training of the ligand-conditioned generator. The two target-specific variants share the explicit generation and CAR formulations. Please refer to Appendix~\ref{app:codes-training} for more details about training.
\end{CODesignKeepParagraph}
\begin{table}[t]
\vspace{5pt}
    \centering
    \footnotesize
    \caption{
    Results of protein-conditioned protein binder design benchmark.
    scRMSD is reported as the mean value / percentage of samples with scRMSD $\leq 2$~\AA.
    Best and second-best results are highlighted in \textbf{bold} and \underline{underlined}, respectively.
    Complete results with detailed standard deviations are provided in Appendix~\ref{app:protein_binder_results}.
    }
    \label{tab:protein_binder_benchmark}

    \setlength{\tabcolsep}{0.3pt}
    \renewcommand{\arraystretch}{1.08}

    \begin{tabular*}{\linewidth}{
        @{\extracolsep{\fill}}
        l
        c c c
        c @{\hspace{2.2pt}} c @{\hspace{2.2pt}} c
        c c
        @{}
    }
        \toprule
        Model
        & \multicolumn{3}{c}{\# Unique Successes $\uparrow$}
        & \multicolumn{3}{c}{scRMSD $\downarrow$ / $\leq 2$~\AA{} (\%) $\uparrow$}
        & Diversity $\uparrow$
        & Novelty $\downarrow$ \\
        \cmidrule(lr){2-4}
        \cmidrule(lr){5-7}

        & Self & MPNN@1 & MPNN@8
        & Self & MPNN@1 & MPNN@8
        & & \\

        \midrule
        \multicolumn{9}{@{}l}{\textit{two-stage}} \\

        RFDiffusion3
        & -- & 6.46 & 10.96
        & -- & 7.24 / 36.12 & 4.17 / 53.02
        & 25.66 & 0.91 \\

        PXDesign
        & -- & 4.59 & 7.01
        & -- & \textbf{5.21} / \textbf{51.15} & \textbf{2.29} / \textbf{74.54}
        & 15.39 & 0.92 \\


        Protpardelle-1c
        & -- & 0.28 & 0.92
        & -- & 16.44 / 10.26 & 12.61 / 20.40
        & \textbf{75.15} & 0.88 \\

        \midrule
        \multicolumn{9}{@{}l}{\textit{co-design}} \\

        Complexa
        & \underline{8.54} & \underline{8.53} & \underline{14.65}
        & \underline{7.91} / \underline{35.06} & 7.50 / 35.28 & 4.49 / 51.18
        & 68.92 & \underline{0.88} \\

        \textbf{CODesign}
        & \textbf{9.84} & \textbf{10.32} & \textbf{18.81}
        & \textbf{7.01} / \textbf{41.64} & \underline{7.01} / \underline{40.53} & \underline{4.09} / \underline{57.16}
        & \underline{73.92} & \textbf{0.87} \\

        \bottomrule
    \end{tabular*}
\vspace{5pt}
\end{table}

\section{Experiments}
\label{sec:experiments}

\subsection{Protein-conditioned Protein Binder Design}
\label{sec:protein_binder_design}

\paragraph{Experimental Configuration}
We follow the protein-target binder design benchmark of Proteína-Complexa \citep{didi2026scaling}, which includes 19 targets from AlphaProteo \citep{zambaldi2024novo} and BindCraft \citep{pacesa2025one}. 
Both self-generated sequences (\textbf{Self}) and the ones redesigned with ProteinMPNN \citep{dauparas2022robust} are evaluated. For each generated backbone, \textbf{MPNN@1} and \textbf{MPNN@8} redesign one and eight binder sequences respectively.
Each sequence is co-folded with its target using AlphaFold2-Multimer \citep{evans2021protein} with templating.
All experiments are performed 5 times with different random seeds.

\paragraph{Metrics}
We calculate average \textbf{Unique Successes} over the 19 targets using Foldseek \citep{van2024fast} for clustering. A design is considered successful if its refolded complex has ipAE $<$ 7~\AA, complex pLDDT $>$ 0.9, and binder $C_\alpha$ scRMSD $<$ 1.5~\AA.
To evaluate the overall consistency, we also calculate the average complex $C_\alpha$ \textbf{scRMSD} and the percentage of samples with scRMSD $\leq 2$~\AA{} for each sequence type. We measure \textbf{Diversity} as the average number of Foldseek clusters among all generated binder structures on each target. \textbf{Novelty} is measured by the mean best-match TM-score against PDB for successful binders, averaged across targets. 

\paragraph{Baselines}
We compare CODesign against two-stage methods including RFDiffusion3 \citep{butcher2025novo}, PXDesign \citep{team2025pxdesign} and Protpardelle-1c \citep{lu2025conditional}; as well as a co-design method Proteína-Complexa \citep{didi2026scaling}. The two-stage models only generate binder structures and use ProteinMPNN for sequence design. All methods are evaluated under the same benchmark protocol without external reward-guided search or hallucination-based refinement. 

\paragraph{Results.}
Table~\ref{tab:protein_binder_benchmark} summarizes performance on protein-target binder design. CODesign achieves the highest number of unique successes across self-generated sequences and both ProteinMPNN redesign settings. It also obtains the strongest self-consistency among co-design methods using model-generated sequences. Although PXDesign attains the highest fraction of designs with $\mathrm{scRMSD}\leq2$~\AA{} after ProteinMPNN redesign, this improvement is accompanied by substantially lower diversity and novelty. Overall, these results show that our consistency-aware model provides a stronger balance between structural consistency, diversity, and novelty than competing methods.

\subsection{Ligand-conditioned Protein Binder Design}
\label{sec:ligand_binder_design}

\paragraph{Experimental Configuration}
We follow the small-molecule binder design benchmark employed by \citet{didi2026scaling}, which includes four targets: SAM, OQO, FAD, and IAI. For each method and target, we generate 200 binders of 100 residues using the same ligand inputs as \citet{didi2026scaling}. We evaluate co-design methods using their self-generated sequences, while two-stage methods use one sequence designed by LigandMPNN \citep{dauparas2025atomic} for each generated backbone. Each sequence is co-folded with its ligand target using RosettaFold-3 (RF3) \citep{corley2025accelerating}.

\paragraph{Metrics}
Similar to the protein target benchmark, we report \textbf{Unique Successes} for each target and their average across the four targets. A design is considered successful if its refolded complex has min-ipAE $<$ 2~\AA, binder $C_\alpha$ scRMSD $<$ 2~\AA, and binder-aligned ligand scRMSD $<$ 5~\AA. Here, min-ipAE is the minimum entry in the protein-ligand blocks of the RF3-predicted pAE matrix. The ligand scRMSD is computed after aligning the refolded binder to the generated one. We additionally evaluate the \textbf{Diversity} and \textbf{Novelty} following the same protocol in the protein-target benchmark. The per-sample generation \textbf{Time} is recorded on a single GPU with batch size set to 1. Reported values are averaged results over five random seeds.

\paragraph{Baselines}
We select both two-stage methods (RFdiffusionAA \citep{krishna2024generalized} and RFDiffusion3 \citep{butcher2025novo}) and co-design methods (Proteína-Complexa \citep{didi2026scaling}, Pallatom-Ligand \citep{wangpallatom}, and DISCO \citep{rector2026general}) for baseline models. For fair comparison, no external reward-guided search or hallucination-based refinement is applied.

\paragraph{Results}
Table~\ref{tab:ligand_binder_benchmark} summarizes performance on ligand-conditioned binder design. CODesign achieves the highest mean number of unique successes, ranking first on OQO and IAI and second on FAD. It also substantially outperforms the strongest competing co-design method, DISCO, in mean unique successes while requiring considerably less inference time. Although Complexa attains higher diversity and faster generation, it produces substantially fewer successful designs; CODesign maintains the second-highest diversity among co-design methods while offering competitive novelty. Overall, these results demonstrate that CODesign generalizes effectively to small-molecule targets and provides a strong balance among success rate, diversity, novelty, and computational cost.

\begin{table}[t]
\vspace{5pt}
    \centering
    \caption{Results of ligand-conditioned protein binder design benchmark. Best and second-best results are highlighted in \textbf{bold} and \underline{underlined}, respectively.
    Complete results with detailed standard deviations are provided in Appendix~\ref{app:ligand_binder_results}.}
    \label{tab:ligand_binder_benchmark}
    \small
    \setlength{\tabcolsep}{3pt}
    \renewcommand{\arraystretch}{1.12}
    \begin{tabular*}{\linewidth}{@{\extracolsep{\fill}}l*{8}{c}@{}}
        \toprule
        Model
        & \multicolumn{5}{c}{\# Unique Successes $\uparrow$}
        & Diversity $\uparrow$
        & Novelty $\downarrow$
        & Time (s) $\downarrow$ \\
        \cmidrule(lr){2-6}
        & SAM & OQO & FAD & IAI & Mean
        & & & \\
        \midrule

        \multicolumn{9}{@{}l}{\textit{two-stage}} \\
        RFdiffusionAA
        & 2.0 & 3.4 & 1.6 & 5.0 & 3.0
        & \underline{135.65} & \textbf{0.74} & 73.78 \\

        RFDiffusion3
        & \underline{9.4} & \underline{14.8} & 20.8 & 11.0 & 14.0
        & 52.50 & 0.83 & \underline{14.96} \\

        \midrule
        \multicolumn{9}{@{}l}{\textit{co-design}} \\

        Complexa
        & 3.2 & 3.0 & 5.8 & 8.4 & 5.1
        & \textbf{147.70} & \underline{0.76} & \textbf{10.27} \\

        Pallatom-Ligand
        & 0.2 & 4.4 & 2.0 & 3.2 & 2.5
        & 17.55 & 0.87 & 40.44 \\

        DISCO
        & \textbf{10.6} & 14.0 & \textbf{25.4} & \underline{12.8} & \underline{15.7}
        & 60.10 & 0.85 & 181.77 \\

        \textbf{CODesign (ours)}
        & 5.8 & \textbf{37.2} & \underline{21.4} & \textbf{18.6} & \textbf{20.8}
        & 102.35 & 0.82 & 39.90 \\

        \bottomrule
    \end{tabular*}
\vspace{5pt}
\end{table}

\begin{figure}[h]
\centering
\includegraphics[width=\linewidth]{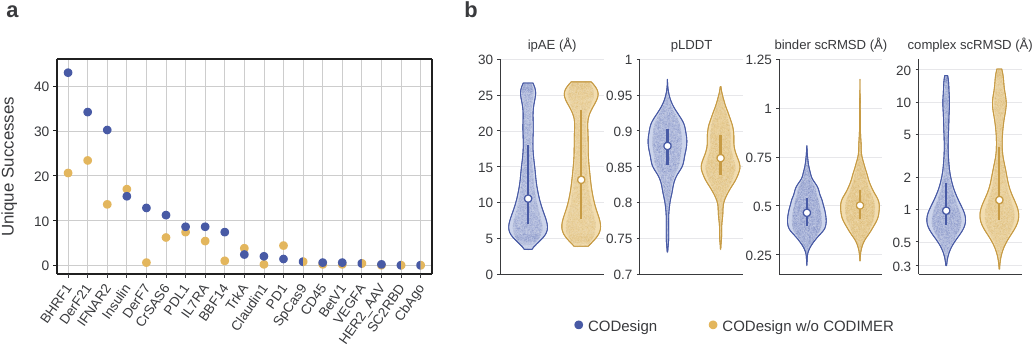}
\caption{Impact of \textsc{Codimer} on the model's performance. \textbf{(a)} Average unique successes across 5 random seeds on each target. Complete results with detailed standard deviations are provided in Appendix~\ref{app:codimer_ablation_results}. \textbf{(b)} Distributions of different confidence and consistency metrics.}
\label{fig:ablation_CODIMER}
\end{figure}

\subsection{Ablation Studies}

\paragraph{Significance of \textsc{Codimer}}

To investigate how \textsc{Codimer} influences the model's performance, we additionally train a CODesign variant only on Teddymer and PDB as in Section~\ref{sec:codes-stages}.
We compare generations without CAR and only the model-generated sequences with 5 random seeds. 

\noindent
\begin{minipage}[t]{0.55\textwidth}
\vspace{0pt}
As demonstrated in Fig.~\ref{fig:ablation_CODIMER}a, the unique successes significantly drop across most targets when training without \textsc{Codimer}. Specifically, including \textsc{Codimer} in training can promote performance across 12/19 targets with a substantial average improvement of 70.9\% (9.46 vs. 5.54) on mean unique successes, as evident in Table~\ref{tab:codes-ablation-summary}.
In Fig.~\ref{fig:ablation_CODIMER}b, we further identify that \textsc{Codimer} increases the performance mainly by improving the physical plausibility of the interface and local structures as well as the complex-level consistency.
\paragraph{Impact of consistency-aware sampling}
To assess the impact of CAR, we evaluate the same CODesign model trained with
\textsc{Codimer}, disabling only CAR during sampling. We evaluate both variants
on the protein-target binder design benchmark over five random seeds.
As shown in Figure~\ref{fig:ablation_CAR}, CAR improves the average number of
unique successes across 19 targets by 4\%, with negligible additional
computational cost. CAR therefore improves consistency by leveraging the
model's intrinsic correction capability, without relying on external rewards.
\end{minipage}
\hfill
\begin{minipage}[t]{0.40\textwidth}
\vspace{0pt}
\centering
\includegraphics[width=0.83\linewidth]
    {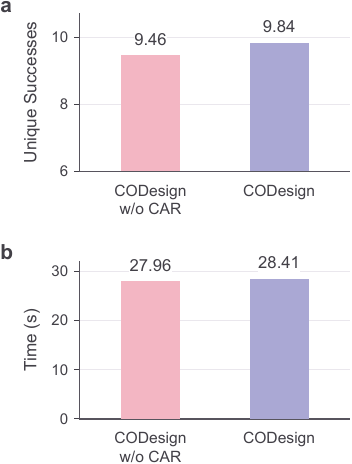}
\captionof{figure}{Impact of CAR on
\textbf{(a)} average unique successes across 19 protein targets and
\textbf{(b)} per-sample generation time.}
\label{fig:ablation_CAR}
\end{minipage}



\section{Conclusions}
We introduced CODesign, a consistency-driven framework for target-conditioned all-atom protein binder co-design. CODesign improves sequence--structure consistency through distilled training pairs and consistency-aware joint resampling, enabling joint generation of compatible sequences, backbones, and local atomic structures. It achieves state-of-the-art in silico success rates on protein- and small-molecule-target benchmarks, with ablations validating both components. These results highlight modality consistency as a key principle for \textit{de novo} protein design. However, evaluation is limited to in silico metrics, and the benchmark's reliance on AlphaFold-based structure prediction and ProteinMPNN-based inverse folding may amplify the apparent importance of consistency. Whether these gains translate into higher binding affinity and experimental success remains to be established through wet-lab validation.

\clearpage
\bibliography{main}
\bibliographystyle{valhalla_references}

\clearpage
\appendix

\section{Details of Codimer}
\label{app:codimer}

\subsection{Data Sources and Preprocessing}
\label{app:codimer-sources}

We collect protein dimers from three sources, requiring both chains to contain 40--300 residues. Source-specific selection proceeds as follows:

\begin{itemize}
    \item \textbf{AFDB high-confidence heterodimers.}
    We use the June 8, 2026 metadata snapshot of the AFDB/NVIDIA high-confidence heterodimer collection \citep{han2026alphafold}. The collection requires maximum ipSAE $\geq 0.6$ and maximum directional pDockQ2 $\geq 0.23$. Retaining entries with available structures and applying the chain-length constraints yields 25,837 dimers.

    \item \textbf{HumanPPI.}
    We use HumanPPI predictions at 80\% expected precision \citep{zhang2025predicting}, retaining protein pairs with available two-chain structural models. When multiple segment-pair models correspond to the same protein pair, we select the model with the highest predicted inter-protein contact probability. Applying the chain-length constraints yields 9,757 dimers.

    \item \textbf{PINDER.}
    We select native holo dimers from the February 2024 release of PINDER \citep{kovtun2024pinder}. We retain heteromeric complexes with biological-interface annotations and distinct entities for the two chains. Both chains must be fully resolved and contain a single connected component. We additionally require no interface atom gaps within 4~\AA{}, at least five intermolecular contacts, and per-chain \texttt{max\_var} metadata values $\leq 0.98$. Together with the chain-length constraints, these filters yield 156,242 dimers.
\end{itemize}

The combined collection contains 191,836 dimers, which serve as the reference structures for consistency distillation.

\subsection{Consistency Distillation}
\label{app:codimer-distillation}

We apply the same construction procedure to all three sources:

\begin{enumerate}
    \item \textbf{Structure preparation.}
    For each input dimer, we use the first coordinate model and retain canonical amino-acid residues with C$\alpha$ coordinates from both chains. We standardize chain identifiers and renumber residues consecutively within each chain. The prepared structure serves as the input to sequence redesign and the reference for subsequent structural comparison.

    \item \textbf{Joint sequence redesign.}
    We use the C$\alpha$-only version of ProteinMPNN \citep{dauparas2022robust} to sample four sequence pairs per dimer at a temperature of 0.1. Both chains are redesigned jointly, rather than fixing the sequence of either interaction partner.

    \item \textbf{All-atom co-folding.}
    Each redesigned sequence pair is independently co-folded with ESMFold2-Fast \citep{candido2026language}, using 10 loops, 68 sampling steps, and one diffusion sample. This produces four candidate complexes per input, or 767,344 candidates in total. Co-folding provides atomistic structures corresponding to the redesigned sequences, including the side-chain geometries and atom compositions associated with the new residue identities \citep{reidenbach2025consistent}.

    \item \textbf{Confidence filtering.}
    We retain candidates with mean pLDDT across both chains $\geq 0.7$ on the $[0,1]$ scale and interface pAE $\leq 10$~\AA{}. Interface contacts are defined by cross-chain C$\alpha$ distances $\leq 8$~\AA{}, and interface pAE is averaged over these residue pairs.

    \item \textbf{Candidate selection.}
    For each input dimer, we select the confidence-qualified candidate with the lowest complex-level C$\alpha$ RMSD to the reference structure. Inputs with no qualifying candidates are discarded. The selected sequence pair and its co-folded all-atom structure constitute one sample in \textit{Codimer}.
\end{enumerate}

This procedure yields 104,528 synthetic sequence--structure pairs. Table~\ref{tab:codimer} summarizes the contribution of each source.

\begin{table}[!htbp]
\centering
\caption{Composition of \textit{Codimer}. Each reference dimer produces four co-folded candidates and contributes at most one sequence--structure pair to the final dataset.}
\label{tab:codimer}
\begin{tabular}{@{}lrr@{}}
\toprule
Source & Reference dimers & Codimer pairs \\
\midrule
AFDB high-confidence heterodimers & 25,837  & 23,823 \\
HumanPPI                         & 9,757   & 5,177  \\
PINDER                           & 156,242 & 75,528 \\
\midrule
Total                            & 191,836 & 104,528 \\
\bottomrule
\end{tabular}
\end{table}


\section{Additional Training Data and Preprocessing}
\label{app:codes-data}

\subsection{Training Datasets}
\label{app:codes-other-data}

In addition to \textsc{Codimer}, CODesign uses Teddymer and PDB structures for protein-conditioned training, and CDDB and PLINDER for ligand adaptation. We describe the role of each dataset below. Their sampling probabilities are reported in Table~\ref{tab:codes-mixtures}.

\paragraph{Teddymer and PDB.}
We use representatives of nonsingleton Teddymer clusters to provide domain--domain interaction examples, following \citet{didi2026scaling}. For experimental data, we use PDB structures selected by the PDB identifier list supplied with the Complexa repository. Entries with multiple protein entities and one chain per entity provide binder--target examples. Entries with repeated copies of an entity, or only one protein entity, provide monomer examples with empty target conditioning. This allows the same generator to learn from both interacting proteins and individual chains. \textsc{Codimer} is added during the second protein-training stage; its construction is described in Appendix~\ref{app:codimer}.

\paragraph{CDDB and PLINDER.}
For ligand adaptation, CDDB provides consistency-distilled monomers of 32--250 residues~\citep{reidenbach2025consistent}. PLINDER provides examples containing one selected protein chain and one connected small-molecule ligand~\citep{durairaj2024plinder}. CDDB examples have no target, whereas PLINDER examples retain the ligand as fixed conditioning information. Mixing the two sources preserves protein sequence--structure supervision while adapting the model to ligand-specific interactions.

\subsection{Cropping and Geometric Preprocessing}
\label{app:codes-cropping}

\paragraph{Protein targets.}
We identify intermolecular contacts using an 8~\AA{} cutoff on cross-chain $C_\alpha$ distances. An interface residue is selected as the binder seed, and a contiguous segment containing this residue is cropped from its chain. The binder contains at most 250 residues. Target residues are selected from a 15~\AA{} neighborhood of the cropped binder, subject to a combined budget of 500 binder and target residues. A valid conditioned crop retains at least 50 target residues and may contain multiple target chains. Examples assigned to monomer generation, or those without a valid target crop, use empty target conditioning. Contact and hotspot annotations are randomly subsampled and dropped during training.

\paragraph{Small-molecule targets.}
We use a spatial crop centered around the protein--ligand system, with a budget of 300 tokens. A protein residue counts as one token and a ligand heavy atom counts as one token. The crop-center distance cutoff is 15~\AA{}. The ligand supplies fixed coordinates and chemical features; only the protein is generated.

\paragraph{Coordinate processing.}
Coordinates are converted to nanometers and augmented by a random rigid rotation. Conditioned examples are centered on the target, while monomers are centered on the protein. Non-$C_\alpha$ coordinates are then expressed relative to the corresponding $C_\alpha$ position. These offsets remain in the shared Cartesian coordinate system; no residue-specific rotation is applied. Protein-training metadata are shuffled with seed 42 and split into 99\% training and 1\% validation examples. The ligand-adaptation configuration uses the supplied CDDB and PLINDER manifests for training without an additional loader-level validation split.

\begin{table}[!htbp]
\centering
\caption{Source-sampling probabilities at each training stage. \textsc{Codimer} is treated as a single distilled dataset. PDB monomers combine repeated-entity and single-entity examples.}
\label{tab:codes-mixtures}
\small
\setlength{\tabcolsep}{5pt}
\begin{tabular*}{\linewidth}{@{\extracolsep{\fill}}lccc@{}}
\toprule
Dataset & Protein I & Protein II & Ligand adaptation \\
\midrule
Teddymer & 0.80 & 0.70 & -- \\
PDB binder--target examples & 0.19 & 0.10 & -- \\
PDB monomers & 0.01 & 0.01 & -- \\
\textsc{Codimer} & -- & 0.19 & -- \\
CDDB monomers & -- & -- & 0.60 \\
PLINDER & -- & -- & 0.40 \\
\bottomrule
\end{tabular*}
\end{table}

\section{Model Architecture}
\label{app:codes-model}

\subsection{Residue and Atom Representations}
\label{app:codes-representation}

We use the notation of Section~\ref{sec:codes-model}: $\mathbf{r}$ denotes the binder sequence, $\mathbf{x}$ its $C_\alpha$ coordinates, $\mathbf{z}$ its non-$C_\alpha$ offsets, and $\mathbf{c}^{\mathrm{target}}$ the fixed target features. The binder contains $n$ residues. For an existing non-$C_\alpha$ atom slot $a$ in residue $i$, the corresponding global coordinate is $\mathbf{x}_i+\mathbf{z}_{ia}$.

Following Prote\'{i}na-Atom\'{i}stica~\citep{reidenbach2025consistent}, the network combines a residue representation with an expanded atom representation. A revealed residue contributes its $C_\alpha$ token and existing non-$C_\alpha$ atoms. A masked residue instead contributes a $C_\alpha$ token and a zero-coordinate pseudo-atom token. Its residue-specific atomic composition is hidden. The pseudo-atom output provides $\mathbf{b}_i^\theta\in\mathbb{R}^{36\times3}$, from which valid atom slots are selected after a residue identity is sampled. This keeps the state dimension fixed while allowing different amino acids to instantiate different atom sets.

\subsection{Target Conditioning and Information Exchange}
\label{app:codes-features}

\paragraph{Input features.}
Binder residue features contain the noisy $C_\alpha$ coordinates and sequence, residue indices, self-conditioning features, and optional clean-backbone or clean-sequence features. Protein targets are represented at the residue level using amino-acid identities, Atom37 coordinates and masks, angular features, and hotspot indicators. Small-molecule targets are represented at the heavy-atom level using coordinates, element and atom-name encodings, charges, and graph Laplacian positional features. Following \citet{didi2026scaling}, target embeddings are concatenated with binder residue embeddings along the token dimension.

Let $n_{\mathrm{target}}$ be the number of target residues for protein conditioning or target heavy atoms for ligand conditioning. The joint token representation has shape $(n+n_{\mathrm{target}})\times d_{\mathrm{res}}$. The pair representation has shape $(n+n_{\mathrm{target}})\times(n+n_{\mathrm{target}})\times d_{\mathrm{pair}}$ and combines binder--binder, target--target, and binder--target features. These features encode sequence separation, chain membership, hotspot annotations, and available geometric distances. Ligand--ligand features additionally encode bond connectivity and bond order.

\paragraph{Denoiser.}
An atom-level Transformer encoder processes local coordinates, atom types, residue identities, and residue indices. Cross-attention passes this information to the residue representation. The joint binder--target representation is then processed by a residue-level Transformer with pair-biased attention. A second cross-attention module passes its updated context to the atom decoder. Pair features remain fixed within each forward evaluation; the network does not use triangular pair updates.

The residue trunk is conditioned on $t_x$ and $t_r$, and the atom branch on $t_z$. Time embeddings modulate attention and transition blocks through adaptive normalization and output scaling. The residue outputs predict $\mathbf{v}_x^\theta$ and $\boldsymbol{\ell}^\theta$, while the atom outputs predict $\mathbf{v}_z^\theta$ and $\mathbf{b}^\theta$. Only binder outputs are retained. The target remains unchanged throughout generation.

\begin{table}[!htbp]
\centering
\caption{CODesign architecture. Protein- and ligand-conditioned variants use the same dimensions and layer counts, with different target feature encoders.}
\label{tab:codes-architecture}
\small
\setlength{\tabcolsep}{5pt}
\begin{tabular*}{\linewidth}{@{\extracolsep{\fill}}lcc@{}}
\toprule
Hyperparameter & Protein-conditioned & Ligand-conditioned \\
\midrule
\multicolumn{3}{@{}l}{\textit{Residue trunk}} \\
Representation dimension $d_{\mathrm{res}}$ & 768 & 768 \\
Conditioning dimension & 256 & 256 \\
Pair dimension $d_{\mathrm{pair}}$ & 256 & 256 \\
Transformer layers & 14 & 14 \\
Attention heads & 12 & 12 \\
Time embedding dimension & 256 & 256 \\
Residue-index embedding dimension & 256 & 256 \\
Residue-type embedding dimension & 512 & 512 \\
Relative sequence-separation dimension & 127 & 127 \\
Noisy / self-conditioned distance bins & 30 / 30 & 30 / 30 \\
Distance range (nm) & 0.1--3.0 & 0.1--3.0 \\
Register tokens & 0 & 0 \\
Triangular pair updates & No & No \\
\midrule
\multicolumn{3}{@{}l}{\textit{Atom encoder--decoder}} \\
Representation / conditioning dimension & 128 / 128 & 128 / 128 \\
Encoder / decoder layers & 2 / 3 & 2 / 3 \\
Self-attention heads & 8 & 8 \\
Cross-attention heads & 8 & 8 \\
Atom-type / residue-type embedding dimension & 128 / 128 & 128 / 128 \\
Residue-index embedding dimension & 128 & 128 \\
\bottomrule
\end{tabular*}
\end{table}

\section{Training Details}
\label{app:codes-training}

\subsection{Stagewise Optimization}
\label{app:codes-optimization}

We first train a protein-conditioned model on Teddymer and PDB, and then continue training after adding \textsc{Codimer}. The first stage initializes shape-compatible parameters from the pretrained Complexa checkpoint. Newly introduced parameters are initialized independently. The second stage starts from the first-stage exponential moving average (EMA) weights. Both stages optimize the full generator. The data mixture changes between stages as shown in Table~\ref{tab:codes-mixtures}, while the architecture and training objectives remain the same.

For small-molecule targets, we initialize from the second-stage protein model and apply LoRA~\citep{hu2022lora} to eligible linear and embedding layers, including the ligand-conditioning projections. For a linear weight matrix, the adapted weight is
\begin{equation}
W=W_0+\frac{\alpha_{\mathrm{LoRA}}}{r_{\mathrm{LoRA}}}BA,
\qquad r_{\mathrm{LoRA}}=32,\quad \alpha_{\mathrm{LoRA}}=64.
\label{eq:codes-lora}
\end{equation}
The base weights and biases are frozen, and only adapter parameters are optimized. Adapter dropout is zero. This stage uses the CDDB--PLINDER mixture in Table~\ref{tab:codes-mixtures}.

All stages use Adam with a learning rate of $10^{-4}$, one optimizer update per batch, and no gradient accumulation. EMA weights are maintained with decay 0.999. Table~\ref{tab:codes-training} reports the training settings and the checkpoint steps selected by the stage-transition and benchmark configurations.

\begin{table}[!htbp]
\centering
\caption{Training settings. Checkpoint steps are optimizer updates within the corresponding stage. Global batch sizes equal the per-GPU batch sizes multiplied by the number of GPUs.}
\label{tab:codes-training}
\small
\setlength{\tabcolsep}{3pt}
\begin{tabular*}{\linewidth}{@{\extracolsep{\fill}}lccc@{}}
\toprule
Setting & Protein I & Protein II & Ligand adaptation \\
\midrule
Number of GPUs & 24 & 64 & 32 \\
Batch size per GPU & 5 & 5 & 3 \\
Global batch size & 120 & 320 & 96 \\
Updates at selected checkpoint & 60,000 & 10,000 & 10,000 \\
Trainable parameters & Full model & Full model & LoRA adapters \\
Optimizer & Adam & Adam & Adam \\
Learning rate & $10^{-4}$ & $10^{-4}$ & $10^{-4}$ \\
Gradient accumulation steps & 1 & 1 & 1 \\
Numerical precision & Float32 & Float32 & Float32 \\
EMA decay & 0.999 & 0.999 & 0.999 \\
Random seed & 42 & 42 & 42 \\
\midrule
Self-conditioning probability & 0.50 & 0.50 & 0.50 \\
Clean-backbone conditioning probability & 0.05 & 0.05 & 0.075 \\
Clean-sequence conditioning probability & 0.05 & 0.05 & 0.075 \\
LoRA rank / scaling coefficient & -- & -- & 32 / 64 \\
\bottomrule
\end{tabular*}
\end{table}

\subsection{Backbone Noise and Modality-Time Sampling}
\label{app:codes-noise}

We follow the conditional paths in Equations~\eqref{eq:codes-flow-x}--\eqref{eq:codes-flow-z}. For the backbone, the initial noise is centered across binder residues and given a random three-dimensional center:
\begin{equation}
\mathbf{x}_{0,i}=\boldsymbol{\epsilon}_{x,i}-\overline{\boldsymbol{\epsilon}}_x+\mathbf{d},
\qquad
\boldsymbol{\epsilon}_{x,i}\sim\mathcal{N}(\mathbf{0},\mathbf{I}_3),
\quad
\mathbf{d}\sim\mathcal{N}(\mathbf{0},c_d^2\mathbf{I}_3),
\quad c_d=0.2.
\label{eq:codes-noise}
\end{equation}
Here, $\overline{\boldsymbol{\epsilon}}_x$ is the mean of the initial Gaussian noise over the binder. Coordinates are in nanometers. Randomizing the noise center prevents the clean binder center from being determined directly from a centered-noise interpolation, encouraging the denoiser to infer placement relative to the target~\citep{didi2026scaling}.

The three times are sampled independently. The coordinate modalities share the same marginal time distribution but do not share their sampled time:
\begin{equation}
\begin{aligned}
t_x,t_z&\overset{\mathrm{ind}}{\sim}
0.02\,\mathcal{U}(0,1)+0.98\,\operatorname{Beta}(1.9,1),\\
t_r&\sim\mathcal{U}(0,1).
\end{aligned}
\label{eq:codes-time-training}
\end{equation}
Independent times expose the network to different combinations of sequence and structure uncertainty. The uniform component in the coordinate distribution retains coverage across noise levels, while the beta component emphasizes more resolved structures.

\subsection{Self-Conditioning and Conditional Training}
\label{app:codes-conditioning}

\begin{CODesignKeepBlock}
With probability 0.5, we perform a preliminary denoiser evaluation without gradient tracking. Its predicted clean $C_\alpha$ coordinates and residue identities are supplied as self-conditioning features in the training evaluation. The preliminary predictions are detached, so the loss does not backpropagate through the first evaluation.

We additionally enable either the clean-backbone feature channel or the clean-sequence feature channel in a subset of training iterations. The two modes are mutually exclusive and equally likely. Their combined probability is 0.1 during protein training and 0.15 during ligand adaptation. They provide ground-truth conditioning features without replacing the independently sampled noisy states or their modality times. Clean-backbone conditioning strengthens structure-to-sequence prediction, while clean-sequence conditioning strengthens sequence-to-structure prediction. These are the same conditional capabilities used by CAR; no separate inverse-folding network is trained.

\end{CODesignKeepBlock}

\subsection{Loss Normalization and Auxiliary Supervision}
\label{app:codes-loss}

The objectives in Section~\ref{sec:codes-model} are evaluated per example and then averaged over the batch. Using the clean-backbone estimate in Equation~\eqref{eq:codes-clean-x}, the backbone loss includes a time-dependent normalization:
\begin{equation}
\mathcal{L}_x=
\frac{1}{n\bigl[(1-t_x)^2+10^{-5}\bigr]}
\sum_{i=1}^{n}\left\|\widehat{\mathbf{x}}_i-\mathbf{x}_i\right\|_2^2.
\label{eq:codes-loss-x-normalized}
\end{equation}
We use the sequence cross-entropy $\mathcal{L}_r$ and local-atom objective $\mathcal{L}_z$ defined in Equations~\eqref{eq:codes-loss-xr} and~\eqref{eq:codes-loss-z}. The local-atom objective supervises $\mathbf{v}_z^\theta$ at revealed residues and $\mathbf{b}^\theta$ at masked residues, using the visibility indicator $V_i$ and atom mask $M_{ia}$ from the main text.

To supervise binder placement, we add a binder--target distance loss. Let $\mathbf{y}_j^{\mathrm{target}}$ denote a target reference point: a $C_\alpha$ coordinate for a protein target, or a heavy-atom coordinate for a small-molecule target. With $n_{\mathrm{target}}$ such points,
\begin{equation}
\mathcal{L}_{\mathrm{bt}}=
\frac{1}{n n_{\mathrm{target}}}
\sum_{i=1}^{n}\sum_{j=1}^{n_{\mathrm{target}}}
\left(
\left\|\widehat{\mathbf{x}}_i-\mathbf{y}_j^{\mathrm{target}}\right\|_2
-
\left\|\mathbf{x}_i-\mathbf{y}_j^{\mathrm{target}}\right\|_2
\right)^2.
\label{eq:codes-loss-bt}
\end{equation}
The term is zero for monomer examples and is enabled only when $t_x>0.3$. The complete training objective is
\begin{equation}
\mathcal{L}_{\mathrm{train}}=
\mathbb{E}\!\left[
\mathcal{L}_x+\mathcal{L}_r+\mathcal{L}_z+
\mathbf{1}\{t_x>0.3\}\mathcal{L}_{\mathrm{bt}}
\right].
\label{eq:codes-training-objective}
\end{equation}
All four terms have unit coefficients. The auxiliary term complements internal reconstruction by directly supervising distances to the conditioning target.

\section{Sampling Details}
\label{app:codes-sampling}

\subsection{Modality Schedules}
\label{app:codes-base-sampler}

We use the same sampling settings for protein and small-molecule targets. A trajectory contains 1,000 global updates. The backbone and sequence schedules finish after 500 updates, while local atomic refinement continues for all 1,000 updates. Let $k\in\{0,\ldots,999\}$ denote the global step and define
\begin{equation}
 f(u)=\frac{1-10^{-2u}}{1-10^{-2}}.
\end{equation}
The modality times are
\begin{equation}
\begin{aligned}
t_x(k)&=f\!\left(\min\{k/500,1\}\right),\\
t_r(k)&=\min\{k/500,1\}^2,\\
t_z(k)&=f\!\left(\min\{k/1000,1\}\right).
\end{aligned}
\label{eq:codes-schedules}
\end{equation}
Each update uses $\Delta t_u=t_u(k+1)-t_u(k)$ for $u\in\{x,r,z\}$. A modality with $\Delta t_u=0$ is not advanced. The exponential backbone schedule provides informative geometry before the quadratic sequence schedule finishes, allowing CAR to revise residue decisions while sequence generation is still active.

\begin{table}[!htbp]
\centering
\caption{Default sampling settings of CODesign. A dash indicates a parameter that does not apply to that modality.}
\label{tab:codes-sampling}
\small
\setlength{\tabcolsep}{4pt}
\begin{tabular*}{\linewidth}{@{\extracolsep{\fill}}lccc@{}}
\toprule
Setting & Backbone $\mathbf{x}$ & Sequence $\mathbf{r}$ & Local atoms $\mathbf{z}$ \\
\midrule
Active updates & 500 & 500 & 1,000 \\
Time schedule & Exponential & Quadratic & Exponential \\
Noise-variance factor & 0.3 & -- & 0.6 \\
Categorical temperature & -- & 0.3 & -- \\
Discrete stochasticity $\eta$ & -- & 5 & -- \\
ODE switching time & 0.98 & -- & 0.98 \\
Final ODE score multiplier & 1.5 & -- & 1.5 \\
\midrule
\multicolumn{4}{@{}l}{\textit{Joint settings}} \\
Total updates & \multicolumn{3}{c}{1,000} \\
Self-conditioning & \multicolumn{3}{c}{Enabled} \\
CAR backbone-time threshold & \multicolumn{3}{c}{$t_x\geq0.95$} \\
CAR check interval & \multicolumn{3}{c}{20 global updates} \\
CAR resampling fraction $\rho$ & \multicolumn{3}{c}{0.4} \\
External reward search / refinement & \multicolumn{3}{c}{Disabled} \\
\bottomrule
\end{tabular*}
\end{table}

\subsection{Coordinate and Sequence Updates}
\label{app:codes-updates}

\paragraph{Continuous coordinates.}
For either coordinate state $\mathbf{y}\in\{\mathbf{x}_{t_x},\mathbf{z}_{t_z}\}$, let $t_y$ be its current time and $\mathbf{v}_y^\theta$ its predicted vector field. The coordinate sampler uses
\begin{equation}
\mathbf{s}_y=\frac{t_y\mathbf{v}_y^\theta-\mathbf{y}}{1-t_y},
\qquad g(t_y)=\frac{1}{t_y+10^{-2}}.
\label{eq:codes-score}
\end{equation}
For $t_y\leq0.98$, we take an Euler--Maruyama step,
\begin{equation}
\mathbf{y}^{+}=\mathbf{y}+
\bigl(\mathbf{v}_y^\theta+g(t_y)\mathbf{s}_y\bigr)\Delta t_y
+\sqrt{2g(t_y)\nu_y\Delta t_y}\,\boldsymbol{\xi},
\qquad\boldsymbol{\xi}\sim\mathcal{N}(\mathbf{0},\mathbf{I}),
\label{eq:codes-continuous-step}
\end{equation}
where $\nu_x=0.3$ and $\nu_z=0.6$. For $t_y>0.98$, we remove the stochastic term and use
\begin{equation}
\mathbf{y}^{+}=\mathbf{y}+
\frac{\mathbf{y}+1.5(1-t_y)\mathbf{s}_y}{t_y}\Delta t_y.
\label{eq:codes-endpoint-step}
\end{equation}
These updates are applied only to active modalities. For local atoms, the resulting coordinate proposal is combined with the sequence transition as described below.

\begin{CODesignKeepBlock}
\paragraph{Discrete sequence.}
We sample amino-acid identities from
$\pi_i(a)=\operatorname{softmax}(\boldsymbol{\ell}_i^\theta/\tau)_a$
with $\tau=0.3$. Following purity-based discrete flow sampling~\citep{campbell2024generative}, we first sample the number of masked positions to reveal:
\begin{equation}
\begin{aligned}
p_{\mathrm{reveal}}&=
\min\!\left\{1,
\frac{1+\eta t_r}{\max(1-t_r,10^{-5})}\Delta t_r\right\},\\
N_{\mathrm{reveal}}&\sim\operatorname{Binomial}(N_{\mathrm{mask}},p_{\mathrm{reveal}}),
\qquad \eta=5.
\end{aligned}
\label{eq:codes-sequence-step}
\end{equation}
Here, $N_{\mathrm{mask}}$ is the current number of masked residues. We select the $N_{\mathrm{reveal}}$ masked positions with the largest $\max_a\pi_i(a)$ and sample their identities categorically. The base sampler also remasks positions with probability $\min\{1,\eta\Delta t_r\}$. Any remaining masks are filled at the final active sequence update. CAR adds targeted corrections to this base sampler rather than replacing its stochastic transitions.

\end{CODesignKeepBlock}

\par\penalty0
\begin{CODesignKeepBlock}
\paragraph{Sequence-dependent atomic states.}
Let $r_i$ and $r_i^{+}$ denote a residue identity immediately before and after a sampling update, and let $M_a(r_i^{+})$ denote whether atom slot $a$ exists for the updated identity. Local atoms follow
\begin{equation}
\mathbf{z}_{ia}^{+}=
\begin{cases}
\mathbf{0},&r_i^{+}=\mathtt{m},\\
M_a(r_i^{+})\mathbf{b}_{ia}^{\theta},
&r_i=\mathtt{m},\ r_i^{+}\ne\mathtt{m},\\
M_a(r_i^{+})\mathbf{z}_{ia}^{\mathrm{prop}},
&r_i\ne\mathtt{m},\ r_i^{+}\ne\mathtt{m},
\end{cases}
\label{eq:codes-atom-transition}
\end{equation}
where $\mathbf{z}^{\mathrm{prop}}$ is the continuous coordinate proposal. Newly revealed residues therefore receive the learned initialization directly. At the end of sampling, existing non-$C_\alpha$ offsets are added to the final $C_\alpha$ coordinates to reconstruct the all-atom binder.

\end{CODesignKeepBlock}

\subsection{Consistency-Aware Joint Resampling}
\label{app:codes-checker}

\paragraph{Assessment context.}
CAR is applied every 20 global updates when $t_x\geq0.95$ and $\Delta t_r>0$. A normal denoiser evaluation first supplies $\widehat{\mathbf{x}}$ and the current sequence prediction. The assessment in Equation~\eqref{eq:codes-cajr-assessment} then masks all sequence identities and hides their non-$C_\alpha$ inputs. Only the assessment sequence time is set to zero; the current noisy $C_\alpha$ coordinates, both coordinate times, and target features are retained. The clean-backbone channel receives $\widehat{\mathbf{x}}$, and the clean-sequence channel is disabled.

The assessment also uses $\widehat{\mathbf{x}}$ for backbone self-conditioning. For sequence self-conditioning, revealed positions retain their current sampled identities, while masked positions retain the normal denoiser's sequence prediction. This provides a structural reassessment without discarding the sequence context of the current trajectory.

\paragraph{Position selection.}
We use the untempered assessment probabilities $q_i(a)$ from Section~\ref{sec:codes-cajr}. The preference gap can be calculated directly from logits:
\begin{equation}
\Delta_i=
\widetilde{\ell}_{i,a_i^\star}^{\theta}
-\widetilde{\ell}_{i,r_{t_r,i}}^{\theta}.
\label{eq:codes-logit-gap}
\end{equation}
This is equivalent to Equation~\eqref{eq:codes-cajr-gap} because the log-softmax normalizer cancels. Define the eligible set
$\mathcal{E}=\{i\in\mathcal{U}:a_i^\star\ne r_{t_r,i}\}$.
The number of selected positions and their selection rule are
\begin{equation}
K_{\mathrm{CAR}}=\min\!\left\{\lceil\rho|\mathcal{U}|\rceil,|\mathcal{E}|\right\},
\qquad
\mathcal{S}=\operatorname{TopK}_{i\in\mathcal{E}}
(\Delta_i,K_{\mathrm{CAR}}),
\qquad\rho=0.4.
\label{eq:codes-budget}
\end{equation}
Thus, the budget is a fraction of revealed positions, and only disagreements are selected.

\paragraph{Correction and continued sampling.}
We apply the joint reset in Equation~\eqref{eq:codes-cajr-reset}. The proposed types replace the corresponding entries in sequence self-conditioning built from the normal prediction. The primary sequence is remasked at those positions, and the associated local atomic states are set to zero. Backbone coordinates are not reset. A fresh denoiser evaluation at the original times then updates all four outputs before the next sampling step. If $\mathcal{S}$ is empty, this reevaluation is skipped.

During ordinary sampling, self-conditioning is taken from the preceding clean prediction. Once the sequence schedule has finished, its self-conditioning is fixed to the finalized sequence. Algorithm~\ref{alg:codes-cajr} summarizes the complete trajectory. With the default schedules, CAR has nine assessment opportunities, at global steps 320, 340, 360, 380, 400, 420, 440, 460, and 480. Each assessment adds one forward evaluation, and a nonempty reset adds one more. The maximum additional cost is therefore 18 denoiser forwards beyond the 1,000 normal forwards. No external folding or inverse-folding model is called.

\begin{algorithm}[tbp]
\caption{CODesign sampling with CAR}
\label{alg:codes-cajr}
\begin{algorithmic}[1]
\Require Denoiser parameters $\theta$, target $\mathbf{c}^{\mathrm{target}}$, binder length $n$, modality schedules
\State Initialize $\mathbf{x}_{t_x}$ from the backbone initial noise, $\mathbf{r}_{t_r}\gets\mathtt{m}^{n}$, and $\mathbf{z}_{t_z}\gets\mathbf{0}$
\State Leave self-conditioning empty for the first denoiser forward
\For{$k=0,\ldots,999$}
    \State Set $(t_x,t_r,t_z)$ and their increments using Equation~\eqref{eq:codes-schedules}
    \State Predict $\mathbf{v}_x^\theta,\boldsymbol{\ell}^\theta,\mathbf{v}_z^\theta,\mathbf{b}^\theta$ with the current states and self-conditioning
    \State Recover $\widehat{\mathbf{x}}$ and predicted residue identities
    \If{$k\bmod20=0$, $t_x\geq0.95$, and $\Delta t_r>0$}
        \State Build the masked assessment and its self-conditioning as in Appendix~\ref{app:codes-checker}
        \State Compute $\widetilde{\boldsymbol{\ell}}^\theta$, $a_i^\star$, and $\Delta_i$ using the same denoiser
        \State Select $\mathcal{S}$ using Equation~\eqref{eq:codes-budget}
        \If{$\mathcal{S}\ne\varnothing$}
            \State Form self-conditioning from the normal prediction and set $r_i^{\mathrm{sc}}\gets a_i^\star$ for $i\in\mathcal{S}$
            \State Reset $r_{t_r,i}\gets\mathtt{m}$ and $\mathbf{z}_{t_z,i}\gets\mathbf{0}$ for $i\in\mathcal{S}$
            \State Keep $\mathbf{x}_{t_x}$, the target, and all modality times unchanged
            \State Rerun the denoiser to refresh all four outputs
        \EndIf
    \EndIf
    \State Store the latest clean backbone and sequence predictions for the next self-conditioning input
    \State Advance active backbone and sequence states according to Appendix~\ref{app:codes-updates}
    \State Update local atoms using Equation~\eqref{eq:codes-atom-transition}
    \State If sequence generation has finished, fix sequence self-conditioning to the final sampled identities
\EndFor
\State \Return Final sequence and existing all-atom coordinates $\mathbf{x}_i$ and $\mathbf{x}_i+\mathbf{z}_{ia}$
\end{algorithmic}
\end{algorithm}

\section{Supplementary Experimental Results}
\label{app:codes-results}

\subsection{Reporting Conventions}
\label{app:codes-reporting}

\begin{CODesignKeepBlock}
The following tables supplement the benchmarks in Sections~\ref{sec:protein_binder_design} and~\ref{sec:ligand_binder_design}, and the ablation studies in the main text. We use the evaluation protocols and success criteria defined in the main text. Each benchmark run contains 200 generated binders per target, and the reported quality metrics are summarized across five independent random seeds.

For a metric $m_{s,j}$ from seed $s$ on target $j$, per-target entries report the mean and standard deviation across seeds. For an aggregate over targets, we first compute the target average within each seed and then summarize those five averages. This preserves the across-seed variation of the aggregate; it is not obtained by averaging the per-target standard deviations. Generation times use batch size one on a single GPU.

\end{CODesignKeepBlock}

\subsection{Protein-Conditioned Binder Design}
\label{app:protein_binder_results}

Table~\ref{tab:protein_summary_details} expands Table~\ref{tab:protein_binder_benchmark} with standard deviations for unique successes and structural consistency. To keep the statistics readable, complex scRMSD and the percentage of designs with scRMSD $\leq2$~\AA{} are shown in separate columns. Tables~\ref{tab:protein_unique_self}--\ref{tab:protein_unique_mpnn8} give per-target unique successes. Tables~\ref{tab:protein_scrmsd_self}--\ref{tab:protein_fraction_mpnn8} report per-target consistency using the same metric definitions.

\begin{table}[htbp]
\centering
\caption{Protein-target benchmark: aggregate unique successes and complex $C_\alpha$ scRMSD. Entries are means $\pm$ standard deviations across five seeds; the threshold column is in percent.}
\label{tab:protein_summary_details}
\footnotesize
\setlength{\tabcolsep}{2pt}
\renewcommand{\arraystretch}{1.08}
\begin{tabular*}{\linewidth}{@{\extracolsep{\fill}}lcccc@{}}
\toprule
Model & Sequence & Unique successes $\uparrow$ & scRMSD (\AA) $\downarrow$ & $\leq2$~\AA{} (\%) $\uparrow$ \\
\midrule
RFDiffusion3 & MPNN@1 & $6.46\,\pm\,0.43$ & $7.24\,\pm\,0.11$ & $36.12\,\pm\,0.95$ \\
RFDiffusion3 & MPNN@8 & $10.96\,\pm\,0.31$ & $4.17\,\pm\,0.04$ & $53.02\,\pm\,0.93$ \\
\midrule
PXDesign & MPNN@1 & $4.59\,\pm\,0.25$ & $5.21\,\pm\,0.10$ & $51.15\,\pm\,0.72$ \\
PXDesign & MPNN@8 & $7.01\,\pm\,0.51$ & $2.29\,\pm\,0.04$ & $74.54\,\pm\,0.47$ \\
\midrule
Protpardelle-1c & MPNN@1 & $0.28\,\pm\,0.10$ & $16.44\,\pm\,0.29$ & $10.26\,\pm\,0.50$ \\
Protpardelle-1c & MPNN@8 & $0.92\,\pm\,0.24$ & $12.61\,\pm\,0.21$ & $20.40\,\pm\,0.53$ \\
\midrule
Complexa & Self & $8.54\,\pm\,0.14$ & $7.91\,\pm\,0.06$ & $35.06\,\pm\,0.53$ \\
Complexa & MPNN@1 & $8.53\,\pm\,0.26$ & $7.50\,\pm\,0.11$ & $35.28\,\pm\,0.75$ \\
Complexa & MPNN@8 & $14.65\,\pm\,0.51$ & $4.49\,\pm\,0.04$ & $51.18\,\pm\,0.41$ \\
\midrule
CODesign & Self & $9.84\,\pm\,0.25$ & $7.01\,\pm\,0.05$ & $41.64\,\pm\,0.55$ \\
CODesign & MPNN@1 & $10.32\,\pm\,0.97$ & $7.01\,\pm\,0.12$ & $40.53\,\pm\,0.71$ \\
CODesign & MPNN@8 & $18.81\,\pm\,0.91$ & $4.09\,\pm\,0.07$ & $57.16\,\pm\,0.64$ \\
\bottomrule
\end{tabular*}
\end{table}

\begin{table}[htbp]
\centering
\caption{Per-target unique successes with Self sequences. Entries report means $\pm$ standard deviations across five seeds.}
\label{tab:protein_unique_self}
\footnotesize
\setlength{\tabcolsep}{2pt}
\renewcommand{\arraystretch}{1.08}
\begin{tabular*}{\linewidth}{@{\extracolsep{\fill}}lcc@{}}
\toprule
Target & Complexa & CODesign \\
\midrule
IFNAR2 & $35.00\,\pm\,3.61$ & $30.00\,\pm\,4.30$ \\
BHRF1 & $22.00\,\pm\,2.65$ & $43.60\,\pm\,6.15$ \\
BBF14 & $12.80\,\pm\,1.64$ & $8.00\,\pm\,1.58$ \\
DerF21 & $20.60\,\pm\,2.88$ & $36.40\,\pm\,4.51$ \\
TrkA & $10.00\,\pm\,1.73$ & $2.60\,\pm\,1.34$ \\
PD1 & $13.20\,\pm\,3.96$ & $2.00\,\pm\,0.00$ \\
Insulin & $12.20\,\pm\,1.48$ & $16.40\,\pm\,3.91$ \\
DerF7 & $10.60\,\pm\,2.30$ & $12.40\,\pm\,3.78$ \\
PDL1 & $7.40\,\pm\,1.52$ & $8.40\,\pm\,2.51$ \\
IL7RA & $4.60\,\pm\,1.14$ & $7.20\,\pm\,2.05$ \\
CrSAS6 & $6.80\,\pm\,1.48$ & $14.20\,\pm\,2.17$ \\
Claudin1 & $1.80\,\pm\,0.45$ & $2.80\,\pm\,2.77$ \\
VEGFA & $1.60\,\pm\,1.14$ & $0.60\,\pm\,0.89$ \\
SpCas9 & $2.40\,\pm\,1.14$ & $1.80\,\pm\,1.10$ \\
SC2RBD & $0.00\,\pm\,0.00$ & $0.00\,\pm\,0.00$ \\
CbAgo & $0.00\,\pm\,0.00$ & $0.00\,\pm\,0.00$ \\
CD45 & $1.00\,\pm\,1.00$ & $0.40\,\pm\,0.55$ \\
BetV1 & $0.00\,\pm\,0.00$ & $0.20\,\pm\,0.45$ \\
HER2\_AAV & $0.20\,\pm\,0.45$ & $0.00\,\pm\,0.00$ \\
\midrule
Mean & $8.54\,\pm\,0.14$ & $9.84\,\pm\,0.25$ \\
\bottomrule
\end{tabular*}
\end{table}

\begin{table}[htbp]
\centering
\caption{Per-target unique successes with MPNN@1 sequences. Entries report means $\pm$ standard deviations across five seeds.}
\label{tab:protein_unique_mpnn1}
\footnotesize
\setlength{\tabcolsep}{2pt}
\renewcommand{\arraystretch}{1.08}
\begin{tabular*}{\linewidth}{@{\extracolsep{\fill}}lccccc@{}}
\toprule
Target & RFDiffusion3 & PXDesign & Protpardelle-1c & Complexa & CODesign \\
\midrule
IFNAR2 & $24.20\,\pm\,3.96$ & $11.80\,\pm\,2.95$ & $0.80\,\pm\,0.84$ & $39.20\,\pm\,3.49$ & $41.20\,\pm\,2.39$ \\
BHRF1 & $9.20\,\pm\,1.10$ & $5.00\,\pm\,1.00$ & $1.20\,\pm\,0.84$ & $19.20\,\pm\,1.64$ & $40.60\,\pm\,9.99$ \\
BBF14 & $1.80\,\pm\,1.79$ & $3.60\,\pm\,1.82$ & $0.00\,\pm\,0.00$ & $10.40\,\pm\,2.70$ & $10.40\,\pm\,1.52$ \\
DerF21 & $1.80\,\pm\,0.45$ & $8.00\,\pm\,1.58$ & $2.60\,\pm\,1.14$ & $23.60\,\pm\,1.82$ & $42.80\,\pm\,5.81$ \\
TrkA & $10.20\,\pm\,1.79$ & $7.80\,\pm\,0.84$ & $0.20\,\pm\,0.45$ & $12.80\,\pm\,2.39$ & $2.40\,\pm\,2.07$ \\
PD1 & $10.60\,\pm\,2.41$ & $6.00\,\pm\,1.87$ & $0.00\,\pm\,0.00$ & $13.80\,\pm\,1.79$ & $2.20\,\pm\,1.64$ \\
Insulin & $4.20\,\pm\,0.45$ & $3.80\,\pm\,1.30$ & $0.00\,\pm\,0.00$ & $9.60\,\pm\,1.14$ & $9.00\,\pm\,1.87$ \\
DerF7 & $4.80\,\pm\,1.10$ & $9.40\,\pm\,3.51$ & $0.00\,\pm\,0.00$ & $12.00\,\pm\,3.00$ & $16.40\,\pm\,4.51$ \\
PDL1 & $9.60\,\pm\,2.97$ & $5.60\,\pm\,2.70$ & $0.00\,\pm\,0.00$ & $8.80\,\pm\,2.39$ & $7.00\,\pm\,2.55$ \\
IL7RA & $6.40\,\pm\,2.41$ & $4.40\,\pm\,1.14$ & $0.00\,\pm\,0.00$ & $2.20\,\pm\,1.30$ & $7.60\,\pm\,2.70$ \\
CrSAS6 & $22.00\,\pm\,1.41$ & $6.80\,\pm\,2.28$ & $0.00\,\pm\,0.00$ & $5.20\,\pm\,3.19$ & $12.80\,\pm\,1.92$ \\
Claudin1 & $5.60\,\pm\,2.07$ & $4.80\,\pm\,1.30$ & $0.00\,\pm\,0.00$ & $1.80\,\pm\,1.30$ & $2.80\,\pm\,2.17$ \\
VEGFA & $1.60\,\pm\,0.55$ & $3.00\,\pm\,0.71$ & $0.00\,\pm\,0.00$ & $0.60\,\pm\,0.89$ & $0.00\,\pm\,0.00$ \\
SpCas9 & $5.40\,\pm\,1.67$ & $1.40\,\pm\,0.55$ & $0.00\,\pm\,0.00$ & $2.40\,\pm\,0.89$ & $0.60\,\pm\,0.89$ \\
SC2RBD & $3.20\,\pm\,1.48$ & $2.60\,\pm\,1.52$ & $0.00\,\pm\,0.00$ & $0.00\,\pm\,0.00$ & $0.00\,\pm\,0.00$ \\
CbAgo & $0.60\,\pm\,0.89$ & $0.00\,\pm\,0.00$ & $0.00\,\pm\,0.00$ & $0.00\,\pm\,0.00$ & $0.00\,\pm\,0.00$ \\
CD45 & $0.00\,\pm\,0.00$ & $0.00\,\pm\,0.00$ & $0.00\,\pm\,0.00$ & $0.00\,\pm\,0.00$ & $0.00\,\pm\,0.00$ \\
BetV1 & $1.60\,\pm\,1.14$ & $2.80\,\pm\,1.92$ & $0.20\,\pm\,0.45$ & $0.20\,\pm\,0.45$ & $0.00\,\pm\,0.00$ \\
HER2\_AAV & $0.00\,\pm\,0.00$ & $0.40\,\pm\,0.55$ & $0.40\,\pm\,0.55$ & $0.20\,\pm\,0.45$ & $0.20\,\pm\,0.45$ \\
\midrule
Mean & $6.46\,\pm\,0.43$ & $4.59\,\pm\,0.25$ & $0.28\,\pm\,0.10$ & $8.53\,\pm\,0.26$ & $10.32\,\pm\,0.97$ \\
\bottomrule
\end{tabular*}
\end{table}

\begin{table}[htbp]
\centering
\caption{Per-target unique successes with MPNN@8 sequences. Entries report means $\pm$ standard deviations across five seeds.}
\label{tab:protein_unique_mpnn8}
\footnotesize
\setlength{\tabcolsep}{2pt}
\renewcommand{\arraystretch}{1.08}
\begin{tabular*}{\linewidth}{@{\extracolsep{\fill}}lccccc@{}}
\toprule
Target & RFDiffusion3 & PXDesign & Protpardelle-1c & Complexa & CODesign \\
\midrule
IFNAR2 & $35.60\,\pm\,4.16$ & $19.00\,\pm\,2.92$ & $2.60\,\pm\,1.14$ & $58.80\,\pm\,2.77$ & $62.60\,\pm\,4.39$ \\
BHRF1 & $11.60\,\pm\,0.55$ & $7.20\,\pm\,1.92$ & $4.80\,\pm\,1.10$ & $23.00\,\pm\,2.55$ & $67.60\,\pm\,6.50$ \\
BBF14 & $4.80\,\pm\,1.30$ & $7.40\,\pm\,2.07$ & $0.20\,\pm\,0.45$ & $19.20\,\pm\,4.09$ & $19.80\,\pm\,3.42$ \\
DerF21 & $5.20\,\pm\,0.84$ & $11.40\,\pm\,4.67$ & $3.40\,\pm\,1.14$ & $34.80\,\pm\,5.45$ & $63.00\,\pm\,5.00$ \\
TrkA & $14.40\,\pm\,2.88$ & $11.20\,\pm\,2.49$ & $0.80\,\pm\,1.30$ & $20.20\,\pm\,1.64$ & $6.00\,\pm\,2.45$ \\
PD1 & $17.40\,\pm\,2.30$ & $8.20\,\pm\,1.10$ & $0.40\,\pm\,0.89$ & $26.20\,\pm\,2.95$ & $4.00\,\pm\,1.41$ \\
Insulin & $7.40\,\pm\,1.95$ & $7.00\,\pm\,1.58$ & $0.40\,\pm\,0.55$ & $20.40\,\pm\,3.21$ & $23.40\,\pm\,5.13$ \\
DerF7 & $9.40\,\pm\,2.07$ & $12.60\,\pm\,2.07$ & $0.40\,\pm\,0.55$ & $25.00\,\pm\,4.69$ & $35.00\,\pm\,4.12$ \\
PDL1 & $17.80\,\pm\,3.27$ & $7.80\,\pm\,0.45$ & $1.80\,\pm\,1.64$ & $17.80\,\pm\,3.83$ & $14.20\,\pm\,4.02$ \\
IL7RA & $10.40\,\pm\,2.19$ & $6.40\,\pm\,2.07$ & $0.00\,\pm\,0.00$ & $4.80\,\pm\,2.28$ & $16.20\,\pm\,2.49$ \\
CrSAS6 & $37.80\,\pm\,2.28$ & $8.20\,\pm\,1.92$ & $0.20\,\pm\,0.45$ & $12.80\,\pm\,1.79$ & $27.20\,\pm\,4.82$ \\
Claudin1 & $8.80\,\pm\,1.64$ & $6.20\,\pm\,0.45$ & $0.60\,\pm\,0.89$ & $5.40\,\pm\,1.95$ & $10.60\,\pm\,1.95$ \\
VEGFA & $5.40\,\pm\,0.89$ & $5.40\,\pm\,1.14$ & $0.00\,\pm\,0.00$ & $3.60\,\pm\,0.55$ & $1.20\,\pm\,1.10$ \\
SpCas9 & $10.80\,\pm\,1.48$ & $2.40\,\pm\,0.55$ & $0.60\,\pm\,0.55$ & $4.60\,\pm\,1.67$ & $3.80\,\pm\,1.48$ \\
SC2RBD & $7.00\,\pm\,1.41$ & $6.00\,\pm\,1.00$ & $0.20\,\pm\,0.45$ & $0.20\,\pm\,0.45$ & $0.20\,\pm\,0.45$ \\
CbAgo & $0.60\,\pm\,0.89$ & $0.00\,\pm\,0.00$ & $0.00\,\pm\,0.00$ & $0.00\,\pm\,0.00$ & $0.00\,\pm\,0.00$ \\
CD45 & $0.40\,\pm\,0.89$ & $0.40\,\pm\,0.55$ & $0.00\,\pm\,0.00$ & $0.00\,\pm\,0.00$ & $0.20\,\pm\,0.45$ \\
BetV1 & $3.20\,\pm\,1.48$ & $4.00\,\pm\,1.41$ & $0.60\,\pm\,0.55$ & $0.60\,\pm\,0.89$ & $2.00\,\pm\,1.00$ \\
HER2\_AAV & $0.20\,\pm\,0.45$ & $2.40\,\pm\,1.14$ & $0.40\,\pm\,0.55$ & $1.00\,\pm\,0.71$ & $0.40\,\pm\,0.55$ \\
\midrule
Mean & $10.96\,\pm\,0.31$ & $7.01\,\pm\,0.51$ & $0.92\,\pm\,0.24$ & $14.65\,\pm\,0.51$ & $18.81\,\pm\,0.91$ \\
\bottomrule
\end{tabular*}
\end{table}

\paragraph{Complex-level consistency.}
The following tables report consistency for all generated binders, not only successful designs. Mean scRMSD and the fraction satisfying the 2~\AA{} threshold are separated to avoid compressing four statistics into one cell. Both quantities are reported for each sequence type.

\begin{table}[htbp]
\centering
\caption{Per-target mean complex $C_\alpha$ scRMSD with Self sequences. Values are in \AA{}; lower is better. Entries report means $\pm$ standard deviations across five seeds.}
\label{tab:protein_scrmsd_self}
\footnotesize
\setlength{\tabcolsep}{2pt}
\renewcommand{\arraystretch}{1.08}
\begin{tabular*}{\linewidth}{@{\extracolsep{\fill}}lcc@{}}
\toprule
Target & Complexa & CODesign \\
\midrule
IFNAR2 & $4.26\,\pm\,0.25$ & $2.17\,\pm\,0.24$ \\
BHRF1 & $1.44\,\pm\,0.16$ & $0.90\,\pm\,0.08$ \\
BBF14 & $9.36\,\pm\,0.42$ & $8.71\,\pm\,0.48$ \\
DerF21 & $3.53\,\pm\,0.14$ & $4.26\,\pm\,0.24$ \\
TrkA & $4.42\,\pm\,0.20$ & $4.73\,\pm\,0.35$ \\
PD1 & $5.08\,\pm\,0.37$ & $4.66\,\pm\,0.50$ \\
Insulin & $3.58\,\pm\,0.23$ & $3.64\,\pm\,0.07$ \\
DerF7 & $4.67\,\pm\,0.27$ & $4.61\,\pm\,0.33$ \\
PDL1 & $3.83\,\pm\,0.43$ & $3.81\,\pm\,0.24$ \\
IL7RA & $6.34\,\pm\,0.33$ & $4.46\,\pm\,0.16$ \\
CrSAS6 & $10.43\,\pm\,0.26$ & $8.59\,\pm\,0.50$ \\
Claudin1 & $5.18\,\pm\,0.34$ & $4.03\,\pm\,0.20$ \\
VEGFA & $8.86\,\pm\,0.41$ & $5.64\,\pm\,0.41$ \\
SpCas9 & $9.48\,\pm\,0.27$ & $10.27\,\pm\,0.81$ \\
SC2RBD & $10.90\,\pm\,0.10$ & $13.97\,\pm\,0.23$ \\
CbAgo & $14.35\,\pm\,0.19$ & $12.11\,\pm\,0.18$ \\
CD45 & $12.53\,\pm\,0.31$ & $10.39\,\pm\,0.45$ \\
BetV1 & $19.27\,\pm\,0.36$ & $16.15\,\pm\,0.34$ \\
HER2\_AAV & $12.77\,\pm\,0.40$ & $10.21\,\pm\,0.11$ \\
\midrule
Mean & $7.91\,\pm\,0.06$ & $7.01\,\pm\,0.05$ \\
\bottomrule
\end{tabular*}
\end{table}

\begin{table}[htbp]
\centering
\caption{Per-target mean complex $C_\alpha$ scRMSD with MPNN@1 sequences. Values are in \AA{}; lower is better. Entries report means $\pm$ standard deviations across five seeds.}
\label{tab:protein_scrmsd_mpnn1}
\footnotesize
\setlength{\tabcolsep}{2pt}
\renewcommand{\arraystretch}{1.08}
\begin{tabular*}{\linewidth}{@{\extracolsep{\fill}}lccccc@{}}
\toprule
Target & RFDiffusion3 & PXDesign & Protpardelle-1c & Complexa & CODesign \\
\midrule
IFNAR2 & $2.56\,\pm\,0.21$ & $4.72\,\pm\,0.36$ & $5.39\,\pm\,0.34$ & $4.50\,\pm\,0.49$ & $2.17\,\pm\,0.21$ \\
BHRF1 & $2.63\,\pm\,0.29$ & $2.18\,\pm\,0.19$ & $7.82\,\pm\,0.45$ & $1.22\,\pm\,0.15$ & $0.87\,\pm\,0.10$ \\
BBF14 & $10.16\,\pm\,0.28$ & $7.59\,\pm\,0.44$ & $45.26\,\pm\,2.20$ & $7.37\,\pm\,0.20$ & $6.33\,\pm\,0.37$ \\
DerF21 & $10.32\,\pm\,0.24$ & $3.48\,\pm\,0.11$ & $19.21\,\pm\,2.51$ & $2.85\,\pm\,0.31$ & $4.92\,\pm\,0.40$ \\
TrkA & $2.81\,\pm\,0.23$ & $2.20\,\pm\,0.10$ & $7.80\,\pm\,0.33$ & $3.77\,\pm\,0.27$ & $4.38\,\pm\,0.32$ \\
PD1 & $3.57\,\pm\,0.35$ & $4.46\,\pm\,0.65$ & $7.81\,\pm\,0.15$ & $3.79\,\pm\,0.15$ & $4.66\,\pm\,0.55$ \\
Insulin & $4.86\,\pm\,0.27$ & $3.80\,\pm\,0.28$ & $8.74\,\pm\,0.23$ & $3.57\,\pm\,0.25$ & $4.11\,\pm\,0.27$ \\
DerF7 & $9.24\,\pm\,0.57$ & $2.77\,\pm\,0.26$ & $18.56\,\pm\,0.70$ & $3.95\,\pm\,0.32$ & $4.00\,\pm\,0.40$ \\
PDL1 & $3.00\,\pm\,0.33$ & $1.90\,\pm\,0.29$ & $6.63\,\pm\,0.15$ & $3.77\,\pm\,0.43$ & $3.46\,\pm\,0.18$ \\
IL7RA & $3.88\,\pm\,0.32$ & $2.11\,\pm\,0.22$ & $14.23\,\pm\,0.47$ & $4.61\,\pm\,0.18$ & $4.45\,\pm\,0.27$ \\
CrSAS6 & $4.74\,\pm\,0.32$ & $6.21\,\pm\,0.91$ & $43.36\,\pm\,1.29$ & $11.44\,\pm\,0.40$ & $9.06\,\pm\,0.39$ \\
Claudin1 & $3.09\,\pm\,0.21$ & $6.43\,\pm\,0.34$ & $12.56\,\pm\,0.26$ & $4.71\,\pm\,0.42$ & $3.87\,\pm\,0.24$ \\
VEGFA & $8.34\,\pm\,0.64$ & $5.77\,\pm\,0.37$ & $16.99\,\pm\,1.23$ & $8.92\,\pm\,0.77$ & $6.53\,\pm\,0.44$ \\
SpCas9 & $7.28\,\pm\,0.51$ & $2.73\,\pm\,0.17$ & $9.65\,\pm\,0.42$ & $11.94\,\pm\,0.46$ & $13.62\,\pm\,0.36$ \\
SC2RBD & $6.93\,\pm\,0.57$ & $5.61\,\pm\,0.17$ & $9.89\,\pm\,0.39$ & $9.45\,\pm\,0.12$ & $9.34\,\pm\,0.15$ \\
CbAgo & $14.33\,\pm\,0.29$ & $8.66\,\pm\,0.53$ & $14.78\,\pm\,0.43$ & $15.46\,\pm\,0.36$ & $12.78\,\pm\,0.37$ \\
CD45 & $13.55\,\pm\,0.32$ & $11.89\,\pm\,0.40$ & $26.76\,\pm\,0.90$ & $14.16\,\pm\,0.17$ & $14.54\,\pm\,0.24$ \\
BetV1 & $12.31\,\pm\,0.32$ & $8.95\,\pm\,0.29$ & $18.19\,\pm\,0.64$ & $15.81\,\pm\,0.31$ & $15.37\,\pm\,0.42$ \\
HER2\_AAV & $13.98\,\pm\,0.23$ & $7.62\,\pm\,0.58$ & $18.73\,\pm\,0.46$ & $11.15\,\pm\,0.48$ & $8.70\,\pm\,0.36$ \\
\midrule
Mean & $7.24\,\pm\,0.11$ & $5.21\,\pm\,0.10$ & $16.44\,\pm\,0.29$ & $7.50\,\pm\,0.11$ & $7.01\,\pm\,0.12$ \\
\bottomrule
\end{tabular*}
\end{table}

\begin{table}[htbp]
\centering
\caption{Per-target mean complex $C_\alpha$ scRMSD with MPNN@8 sequences. Values are in \AA{}; lower is better. Entries report means $\pm$ standard deviations across five seeds.}
\label{tab:protein_scrmsd_mpnn8}
\footnotesize
\setlength{\tabcolsep}{2pt}
\renewcommand{\arraystretch}{1.08}
\begin{tabular*}{\linewidth}{@{\extracolsep{\fill}}lccccc@{}}
\toprule
Target & RFDiffusion3 & PXDesign & Protpardelle-1c & Complexa & CODesign \\
\midrule
IFNAR2 & $1.09\,\pm\,0.05$ & $1.62\,\pm\,0.24$ & $2.55\,\pm\,0.18$ & $1.91\,\pm\,0.20$ & $1.06\,\pm\,0.05$ \\
BHRF1 & $1.31\,\pm\,0.13$ & $1.15\,\pm\,0.13$ & $4.83\,\pm\,0.30$ & $0.82\,\pm\,0.03$ & $0.57\,\pm\,0.02$ \\
BBF14 & $5.97\,\pm\,0.30$ & $4.07\,\pm\,0.39$ & $42.49\,\pm\,2.47$ & $3.56\,\pm\,0.16$ & $3.14\,\pm\,0.27$ \\
DerF21 & $7.76\,\pm\,0.22$ & $1.32\,\pm\,0.10$ & $13.80\,\pm\,2.35$ & $1.53\,\pm\,0.08$ & $2.40\,\pm\,0.25$ \\
TrkA & $1.23\,\pm\,0.09$ & $0.86\,\pm\,0.08$ & $4.47\,\pm\,0.13$ & $1.83\,\pm\,0.08$ & $2.25\,\pm\,0.27$ \\
PD1 & $1.47\,\pm\,0.13$ & $1.62\,\pm\,0.11$ & $4.42\,\pm\,0.31$ & $1.58\,\pm\,0.17$ & $1.91\,\pm\,0.17$ \\
Insulin & $1.82\,\pm\,0.18$ & $1.12\,\pm\,0.09$ & $4.52\,\pm\,0.24$ & $1.67\,\pm\,0.09$ & $1.68\,\pm\,0.21$ \\
DerF7 & $4.27\,\pm\,0.17$ & $1.17\,\pm\,0.11$ & $15.07\,\pm\,0.83$ & $2.07\,\pm\,0.20$ & $1.87\,\pm\,0.21$ \\
PDL1 & $1.15\,\pm\,0.09$ & $0.71\,\pm\,0.03$ & $3.73\,\pm\,0.25$ & $1.75\,\pm\,0.23$ & $1.36\,\pm\,0.06$ \\
IL7RA & $1.83\,\pm\,0.09$ & $0.95\,\pm\,0.04$ & $10.60\,\pm\,0.27$ & $2.17\,\pm\,0.16$ & $2.17\,\pm\,0.17$ \\
CrSAS6 & $1.55\,\pm\,0.22$ & $2.01\,\pm\,0.32$ & $40.11\,\pm\,1.22$ & $6.74\,\pm\,0.25$ & $4.55\,\pm\,0.23$ \\
Claudin1 & $1.03\,\pm\,0.05$ & $2.12\,\pm\,0.22$ & $7.25\,\pm\,0.20$ & $1.93\,\pm\,0.14$ & $1.51\,\pm\,0.14$ \\
VEGFA & $2.96\,\pm\,0.29$ & $1.20\,\pm\,0.08$ & $10.14\,\pm\,1.37$ & $3.81\,\pm\,0.21$ & $2.13\,\pm\,0.33$ \\
SpCas9 & $3.30\,\pm\,0.30$ & $1.20\,\pm\,0.04$ & $5.99\,\pm\,0.40$ & $7.07\,\pm\,0.39$ & $8.37\,\pm\,0.05$ \\
SC2RBD & $2.78\,\pm\,0.18$ & $2.57\,\pm\,0.15$ & $5.12\,\pm\,0.22$ & $5.88\,\pm\,0.14$ & $4.99\,\pm\,0.15$ \\
CbAgo & $11.11\,\pm\,0.38$ & $3.59\,\pm\,0.16$ & $11.70\,\pm\,0.37$ & $12.52\,\pm\,0.38$ & $10.08\,\pm\,0.28$ \\
CD45 & $10.07\,\pm\,0.18$ & $8.85\,\pm\,0.25$ & $23.51\,\pm\,0.88$ & $10.89\,\pm\,0.21$ & $11.59\,\pm\,0.29$ \\
BetV1 & $7.98\,\pm\,0.22$ & $4.19\,\pm\,0.32$ & $15.46\,\pm\,0.65$ & $10.84\,\pm\,0.49$ & $11.18\,\pm\,0.53$ \\
HER2\_AAV & $10.61\,\pm\,0.41$ & $3.19\,\pm\,0.16$ & $13.78\,\pm\,0.62$ & $6.71\,\pm\,0.42$ & $4.89\,\pm\,0.34$ \\
\midrule
Mean & $4.17\,\pm\,0.04$ & $2.29\,\pm\,0.04$ & $12.61\,\pm\,0.21$ & $4.49\,\pm\,0.04$ & $4.09\,\pm\,0.07$ \\
\bottomrule
\end{tabular*}
\end{table}

\begin{table}[htbp]
\centering
\caption{Per-target percentage of designs with complex $C_\alpha$ scRMSD $\leq2$~\AA{} with Self sequences. Values are percentages; higher is better. Entries report means $\pm$ standard deviations across five seeds.}
\label{tab:protein_fraction_self}
\footnotesize
\setlength{\tabcolsep}{2pt}
\renewcommand{\arraystretch}{1.08}
\begin{tabular*}{\linewidth}{@{\extracolsep{\fill}}lcc@{}}
\toprule
Target & Complexa & CODesign \\
\midrule
IFNAR2 & $58.10\,\pm\,1.08$ & $79.20\,\pm\,2.95$ \\
BHRF1 & $94.00\,\pm\,1.27$ & $96.90\,\pm\,1.14$ \\
BBF14 & $26.50\,\pm\,2.78$ & $31.10\,\pm\,3.96$ \\
DerF21 & $42.10\,\pm\,2.58$ & $69.20\,\pm\,3.05$ \\
TrkA & $48.70\,\pm\,2.75$ & $34.50\,\pm\,3.48$ \\
PD1 & $45.60\,\pm\,4.36$ & $46.10\,\pm\,6.05$ \\
Insulin & $59.50\,\pm\,2.67$ & $67.00\,\pm\,2.52$ \\
DerF7 & $54.80\,\pm\,2.20$ & $53.10\,\pm\,2.04$ \\
PDL1 & $53.70\,\pm\,4.02$ & $49.00\,\pm\,2.94$ \\
IL7RA & $32.00\,\pm\,4.72$ & $57.80\,\pm\,1.15$ \\
CrSAS6 & $22.30\,\pm\,1.68$ & $33.80\,\pm\,3.05$ \\
Claudin1 & $47.40\,\pm\,3.09$ & $56.40\,\pm\,1.14$ \\
VEGFA & $33.60\,\pm\,1.56$ & $50.40\,\pm\,2.04$ \\
SpCas9 & $28.50\,\pm\,1.87$ & $22.30\,\pm\,4.31$ \\
SC2RBD & $0.50\,\pm\,0.50$ & $0.00\,\pm\,0.00$ \\
CbAgo & $0.20\,\pm\,0.27$ & $0.00\,\pm\,0.00$ \\
CD45 & $1.90\,\pm\,0.74$ & $9.40\,\pm\,2.27$ \\
BetV1 & $0.30\,\pm\,0.27$ & $2.50\,\pm\,1.00$ \\
HER2\_AAV & $16.50\,\pm\,2.60$ & $32.50\,\pm\,1.80$ \\
\midrule
Mean & $35.06\,\pm\,0.53$ & $41.64\,\pm\,0.55$ \\
\bottomrule
\end{tabular*}
\end{table}

\begin{table}[htbp]
\centering
\caption{Per-target percentage of designs with complex $C_\alpha$ scRMSD $\leq2$~\AA{} with MPNN@1 sequences. Values are percentages; higher is better. Entries report means $\pm$ standard deviations across five seeds.}
\label{tab:protein_fraction_mpnn1}
\footnotesize
\setlength{\tabcolsep}{2pt}
\renewcommand{\arraystretch}{1.08}
\begin{tabular*}{\linewidth}{@{\extracolsep{\fill}}lccccc@{}}
\toprule
Target & RFDiffusion3 & PXDesign & Protpardelle-1c & Complexa & CODesign \\
\midrule
IFNAR2 & $80.90\,\pm\,3.19$ & $57.60\,\pm\,2.90$ & $40.30\,\pm\,2.08$ & $57.90\,\pm\,2.41$ & $80.20\,\pm\,2.51$ \\
BHRF1 & $78.80\,\pm\,3.63$ & $83.90\,\pm\,2.22$ & $25.80\,\pm\,3.33$ & $94.60\,\pm\,1.08$ & $97.10\,\pm\,1.47$ \\
BBF14 & $4.80\,\pm\,1.68$ & $20.30\,\pm\,3.07$ & $0.10\,\pm\,0.22$ & $27.30\,\pm\,1.44$ & $37.60\,\pm\,3.17$ \\
DerF21 & $0.80\,\pm\,0.57$ & $68.60\,\pm\,1.56$ & $8.40\,\pm\,2.16$ & $57.50\,\pm\,3.66$ & $66.00\,\pm\,2.32$ \\
TrkA & $65.70\,\pm\,2.51$ & $75.10\,\pm\,2.10$ & $14.90\,\pm\,3.29$ & $54.50\,\pm\,3.71$ & $44.70\,\pm\,2.68$ \\
PD1 & $58.10\,\pm\,2.92$ & $54.10\,\pm\,1.92$ & $15.50\,\pm\,1.62$ & $53.20\,\pm\,3.09$ & $32.20\,\pm\,3.65$ \\
Insulin & $43.10\,\pm\,1.78$ & $60.10\,\pm\,3.34$ & $6.80\,\pm\,1.60$ & $56.50\,\pm\,3.62$ & $59.90\,\pm\,3.44$ \\
DerF7 & $20.50\,\pm\,3.52$ & $73.00\,\pm\,3.86$ & $3.60\,\pm\,1.56$ & $57.50\,\pm\,3.45$ & $61.90\,\pm\,2.97$ \\
PDL1 & $64.80\,\pm\,3.25$ & $81.50\,\pm\,4.40$ & $13.70\,\pm\,2.59$ & $45.60\,\pm\,4.25$ & $50.80\,\pm\,2.17$ \\
IL7RA & $40.50\,\pm\,3.08$ & $73.50\,\pm\,3.16$ & $1.00\,\pm\,0.50$ & $33.80\,\pm\,1.68$ & $53.00\,\pm\,4.51$ \\
CrSAS6 & $63.90\,\pm\,2.66$ & $44.90\,\pm\,6.78$ & $0.50\,\pm\,0.61$ & $16.50\,\pm\,2.26$ & $29.60\,\pm\,3.80$ \\
Claudin1 & $71.80\,\pm\,1.60$ & $47.80\,\pm\,2.97$ & $7.00\,\pm\,1.46$ & $49.30\,\pm\,4.80$ & $60.70\,\pm\,3.70$ \\
VEGFA & $21.20\,\pm\,4.45$ & $50.70\,\pm\,2.80$ & $1.30\,\pm\,0.57$ & $27.80\,\pm\,4.06$ & $42.60\,\pm\,1.92$ \\
SpCas9 & $50.60\,\pm\,3.70$ & $76.00\,\pm\,3.04$ & $33.80\,\pm\,2.49$ & $14.70\,\pm\,1.75$ & $8.10\,\pm\,1.75$ \\
SC2RBD & $17.90\,\pm\,4.55$ & $27.00\,\pm\,1.80$ & $5.30\,\pm\,2.44$ & $1.50\,\pm\,1.41$ & $6.10\,\pm\,1.88$ \\
CbAgo & $1.30\,\pm\,0.76$ & $33.70\,\pm\,4.40$ & $2.00\,\pm\,1.32$ & $0.00\,\pm\,0.00$ & $0.00\,\pm\,0.00$ \\
CD45 & $0.40\,\pm\,0.55$ & $0.80\,\pm\,0.76$ & $6.10\,\pm\,1.29$ & $0.10\,\pm\,0.22$ & $0.10\,\pm\,0.22$ \\
BetV1 & $1.20\,\pm\,0.97$ & $10.80\,\pm\,1.52$ & $2.20\,\pm\,1.04$ & $0.40\,\pm\,0.22$ & $1.90\,\pm\,1.64$ \\
HER2\_AAV & $0.00\,\pm\,0.00$ & $32.50\,\pm\,2.09$ & $6.70\,\pm\,1.35$ & $21.70\,\pm\,4.74$ & $37.60\,\pm\,1.75$ \\
\midrule
Mean & $36.12\,\pm\,0.95$ & $51.15\,\pm\,0.72$ & $10.26\,\pm\,0.50$ & $35.28\,\pm\,0.75$ & $40.53\,\pm\,0.71$ \\
\bottomrule
\end{tabular*}
\end{table}

\begin{table}[htbp]
\centering
\caption{Per-target percentage of designs with complex $C_\alpha$ scRMSD $\leq2$~\AA{} with MPNN@8 sequences. Values are percentages; higher is better. Entries report means $\pm$ standard deviations across five seeds.}
\label{tab:protein_fraction_mpnn8}
\footnotesize
\setlength{\tabcolsep}{2pt}
\renewcommand{\arraystretch}{1.08}
\begin{tabular*}{\linewidth}{@{\extracolsep{\fill}}lccccc@{}}
\toprule
Target & RFDiffusion3 & PXDesign & Protpardelle-1c & Complexa & CODesign \\
\midrule
IFNAR2 & $92.30\,\pm\,0.76$ & $83.80\,\pm\,3.37$ & $65.00\,\pm\,3.59$ & $80.60\,\pm\,2.38$ & $90.00\,\pm\,1.90$ \\
BHRF1 & $88.30\,\pm\,2.33$ & $89.50\,\pm\,2.21$ & $41.20\,\pm\,4.91$ & $96.50\,\pm\,0.50$ & $99.70\,\pm\,0.45$ \\
BBF14 & $13.30\,\pm\,2.84$ & $39.40\,\pm\,6.66$ & $0.30\,\pm\,0.27$ & $46.90\,\pm\,2.51$ & $60.30\,\pm\,3.55$ \\
DerF21 & $2.80\,\pm\,2.08$ & $90.50\,\pm\,1.00$ & $21.70\,\pm\,3.58$ & $81.50\,\pm\,2.47$ & $80.90\,\pm\,2.41$ \\
TrkA & $86.80\,\pm\,2.71$ & $94.90\,\pm\,1.14$ & $30.00\,\pm\,3.43$ & $75.20\,\pm\,2.33$ & $66.50\,\pm\,5.26$ \\
PD1 & $83.20\,\pm\,2.20$ & $84.00\,\pm\,1.22$ & $30.80\,\pm\,3.27$ & $77.50\,\pm\,4.56$ & $63.90\,\pm\,4.17$ \\
Insulin & $73.90\,\pm\,3.23$ & $87.30\,\pm\,2.44$ & $21.90\,\pm\,1.92$ & $76.70\,\pm\,2.08$ & $82.30\,\pm\,2.39$ \\
DerF7 & $50.10\,\pm\,2.04$ & $91.70\,\pm\,2.61$ & $8.10\,\pm\,2.77$ & $73.00\,\pm\,2.42$ & $79.50\,\pm\,3.10$ \\
PDL1 & $87.80\,\pm\,0.91$ & $98.00\,\pm\,0.79$ & $32.90\,\pm\,4.93$ & $72.60\,\pm\,6.77$ & $79.60\,\pm\,2.25$ \\
IL7RA & $71.10\,\pm\,2.30$ & $97.60\,\pm\,1.78$ & $1.90\,\pm\,0.89$ & $62.50\,\pm\,4.14$ & $74.30\,\pm\,0.91$ \\
CrSAS6 & $86.80\,\pm\,3.01$ & $78.10\,\pm\,3.78$ & $1.30\,\pm\,0.91$ & $31.30\,\pm\,3.23$ & $53.40\,\pm\,3.03$ \\
Claudin1 & $93.60\,\pm\,0.89$ & $80.40\,\pm\,3.15$ & $19.90\,\pm\,3.54$ & $74.60\,\pm\,2.77$ & $82.30\,\pm\,3.05$ \\
VEGFA & $45.60\,\pm\,5.89$ & $87.00\,\pm\,1.80$ & $5.00\,\pm\,1.27$ & $50.50\,\pm\,2.94$ & $70.80\,\pm\,4.42$ \\
SpCas9 & $74.30\,\pm\,1.86$ & $94.40\,\pm\,0.82$ & $52.80\,\pm\,2.39$ & $26.70\,\pm\,3.01$ & $18.50\,\pm\,1.84$ \\
SC2RBD & $45.10\,\pm\,4.20$ & $57.60\,\pm\,2.82$ & $18.60\,\pm\,2.75$ & $6.80\,\pm\,1.92$ & $21.10\,\pm\,0.96$ \\
CbAgo & $3.50\,\pm\,2.00$ & $61.40\,\pm\,2.79$ & $7.10\,\pm\,2.43$ & $0.10\,\pm\,0.22$ & $0.20\,\pm\,0.27$ \\
CD45 & $2.50\,\pm\,0.94$ & $3.60\,\pm\,1.02$ & $9.80\,\pm\,1.72$ & $0.60\,\pm\,0.22$ & $1.00\,\pm\,1.17$ \\
BetV1 & $6.10\,\pm\,2.58$ & $32.80\,\pm\,3.21$ & $3.80\,\pm\,2.02$ & $1.90\,\pm\,0.89$ & $8.60\,\pm\,2.86$ \\
HER2\_AAV & $0.20\,\pm\,0.27$ & $64.30\,\pm\,3.65$ & $15.50\,\pm\,2.37$ & $37.00\,\pm\,4.40$ & $53.10\,\pm\,2.77$ \\
\midrule
Mean & $53.02\,\pm\,0.93$ & $74.54\,\pm\,0.47$ & $20.40\,\pm\,0.53$ & $51.18\,\pm\,0.41$ & $57.16\,\pm\,0.64$ \\
\bottomrule
\end{tabular*}
\end{table}

\Needspace{20\baselineskip}
\subsection{Ligand-Conditioned Binder Design}
\label{app:ligand_binder_results}

Table~\ref{tab:ligand_unique_details} provides the per-target unique successes and four-target mean corresponding to Table~\ref{tab:ligand_binder_benchmark}. Co-design methods use their generated sequences, and the two-stage methods use one LigandMPNN sequence per generated backbone, as in Section~\ref{sec:ligand_binder_design}.

\begin{table}[H]
\centering
\caption{Ligand-target unique successes. Entries report means $\pm$ standard deviations across five seeds. The final column averages the four targets within each seed.}
\label{tab:ligand_unique_details}
\footnotesize
\setlength{\tabcolsep}{2pt}
\renewcommand{\arraystretch}{1.08}
\begin{tabular*}{\linewidth}{@{\extracolsep{\fill}}lccccc@{}}
\toprule
Model & SAM & OQO & FAD & IAI & Mean \\
\midrule
RFdiffusionAA & $2.0\,\pm\,1.4$ & $3.4\,\pm\,1.3$ & $1.6\,\pm\,1.1$ & $5.0\,\pm\,2.5$ & $3.0\,\pm\,0.4$ \\
RFDiffusion3 & $9.4\,\pm\,3.2$ & $14.8\,\pm\,2.4$ & $20.8\,\pm\,2.6$ & $11.0\,\pm\,1.2$ & $14.0\,\pm\,1.1$ \\
Complexa & $3.2\,\pm\,1.5$ & $3.0\,\pm\,1.2$ & $5.8\,\pm\,2.9$ & $8.4\,\pm\,1.7$ & $5.1\,\pm\,0.7$ \\
Pallatom-Ligand & $0.2\,\pm\,0.4$ & $4.4\,\pm\,1.1$ & $2.0\,\pm\,1.0$ & $3.2\,\pm\,1.3$ & $2.5\,\pm\,0.5$ \\
DISCO & $10.6\,\pm\,3.1$ & $14.0\,\pm\,2.5$ & $25.4\,\pm\,3.8$ & $12.8\,\pm\,0.8$ & $15.7\,\pm\,1.0$ \\
CODesign & $5.8\,\pm\,1.6$ & $37.2\,\pm\,3.8$ & $21.4\,\pm\,2.3$ & $18.6\,\pm\,2.7$ & $20.8\,\pm\,1.3$ \\
\bottomrule
\end{tabular*}
\end{table}

\subsection{Ablation Studies}
\label{app:codes-ablation-results}
\label{app:codimer_ablation_results}
\label{app:car_ablation_results}

\paragraph{Effect of \textsc{Codimer}.}
We compare the model trained without \textsc{Codimer} with the model trained using \textsc{Codimer}, disabling CAR in both cases. Both variants use the same architecture and training hyperparameters and are evaluated with self-generated sequences on the 19 protein targets. Table~\ref{tab:codes-ablation-summary} gives the aggregate comparison, and Table~\ref{tab:codes-ablation-targets} gives the per-target results corresponding to Figure~\ref{fig:ablation_CODIMER}. For the distributions in that figure, the top 100 values of each metric are selected independently on each target: the lowest values for ipAE and scRMSD, and the highest values for pLDDT. The selected samples can therefore differ between metrics.

\paragraph{Effect of CAR.}
We use the same \textsc{Codimer}-trained checkpoint with CAR enabled or disabled. All other generation and evaluation settings are unchanged. The comparison therefore measures the effect of inference-time consistency correction rather than a change in training. Table~\ref{tab:codes-ablation-summary} also reports the mean unique successes and timing values shown in Figure~\ref{fig:ablation_CAR}. Quality metrics are summarized across five seeds.

\begin{table}[htbp]
\centering
\caption{Controlled ablations on protein targets using Self sequences. Quality entries report means $\pm$ standard deviations across five seeds; times correspond to Figure~\ref{fig:ablation_CAR}. A dash indicates that generation time was not evaluated.}
\label{tab:codes-ablation-summary}
\footnotesize
\setlength{\tabcolsep}{2pt}
\renewcommand{\arraystretch}{1.08}
\begin{tabular*}{\linewidth}{@{\extracolsep{\fill}}lcccc@{}}
\toprule
Variant & \textsc{Codimer} & CAR & Unique successes $\uparrow$ & Time (s) \\
\midrule
Without \textsc{Codimer}, without CAR & No & No & $5.54\,\pm\,0.48$ & -- \\
Without CAR & Yes & No & $9.46\,\pm\,0.65$ & 27.96 \\
CODesign & Yes & Yes & $9.84\,\pm\,0.25$ & 28.41 \\
\bottomrule
\end{tabular*}
\end{table}

\begin{table}[htbp]
\centering
\caption{Per-target unique successes in the \textsc{Codimer} and CAR ablations. All variants use Self sequences. Entries report means $\pm$ standard deviations across five seeds.}
\label{tab:codes-ablation-targets}
\small
\setlength{\tabcolsep}{3pt}
\renewcommand{\arraystretch}{1.08}
\begin{tabular*}{\linewidth}{@{\extracolsep{\fill}}lccc@{}}
\toprule
Target & \shortstack{Without \textsc{Codimer}\\without CAR} & \shortstack{With \textsc{Codimer}\\without CAR} & CODesign \\
\midrule
IFNAR2 & $13.60\,\pm\,1.14$ & $30.20\,\pm\,7.12$ & $30.00\,\pm\,4.30$ \\
BHRF1 & $20.60\,\pm\,6.27$ & $43.00\,\pm\,4.85$ & $43.60\,\pm\,6.15$ \\
BBF14 & $1.00\,\pm\,1.00$ & $7.40\,\pm\,1.52$ & $8.00\,\pm\,1.58$ \\
DerF21 & $23.40\,\pm\,3.91$ & $34.20\,\pm\,3.27$ & $36.40\,\pm\,4.51$ \\
TrkA & $3.80\,\pm\,0.84$ & $2.40\,\pm\,0.55$ & $2.60\,\pm\,1.34$ \\
PD1 & $4.40\,\pm\,1.34$ & $1.40\,\pm\,0.55$ & $2.00\,\pm\,0.00$ \\
Insulin & $17.00\,\pm\,2.00$ & $15.40\,\pm\,1.67$ & $16.40\,\pm\,3.91$ \\
DerF7 & $0.60\,\pm\,0.55$ & $12.80\,\pm\,2.68$ & $12.40\,\pm\,3.78$ \\
PDL1 & $7.40\,\pm\,1.14$ & $8.60\,\pm\,1.14$ & $8.40\,\pm\,2.51$ \\
IL7RA & $5.40\,\pm\,1.14$ & $8.60\,\pm\,1.52$ & $7.20\,\pm\,2.05$ \\
CrSAS6 & $6.20\,\pm\,1.92$ & $11.20\,\pm\,2.59$ & $14.20\,\pm\,2.17$ \\
Claudin1 & $0.20\,\pm\,0.45$ & $2.00\,\pm\,1.58$ & $2.80\,\pm\,2.77$ \\
VEGFA & $0.40\,\pm\,0.55$ & $0.40\,\pm\,0.55$ & $0.60\,\pm\,0.89$ \\
SpCas9 & $0.80\,\pm\,1.30$ & $0.80\,\pm\,0.84$ & $1.80\,\pm\,1.10$ \\
SC2RBD & $0.00\,\pm\,0.00$ & $0.00\,\pm\,0.00$ & $0.00\,\pm\,0.00$ \\
CbAgo & $0.00\,\pm\,0.00$ & $0.00\,\pm\,0.00$ & $0.00\,\pm\,0.00$ \\
CD45 & $0.20\,\pm\,0.45$ & $0.60\,\pm\,1.34$ & $0.40\,\pm\,0.55$ \\
BetV1 & $0.20\,\pm\,0.45$ & $0.60\,\pm\,0.55$ & $0.20\,\pm\,0.45$ \\
HER2\_AAV & $0.00\,\pm\,0.00$ & $0.20\,\pm\,0.45$ & $0.00\,\pm\,0.00$ \\
\midrule
Mean & $5.54\,\pm\,0.48$ & $9.46\,\pm\,0.65$ & $9.84\,\pm\,0.25$ \\
\bottomrule
\end{tabular*}
\end{table}

\end{document}